\documentclass[useAMS,usenatbib,usegraphicx]{mn2e}

\usepackage{times}

\newcommand{\smallcolsize}{0.40}
\newcommand{\singlecolsize}{0.45}
\newcommand{\middlecolsize}{0.80}

\newcommand{\sqdeg}{{\rm deg}^2}
\newcommand{\persqdeg}{{\rm deg}^{-2}}

\newcommand{\dsg}{\Delta_{\rm sg}}
\newcommand{\dsgjk}{\Delta_{\rm sg,jk}}
\newcommand{\effsb}{\mu_{r,50}}
\newcommand{\rscat}{{\cal R}_{\rm s}}

\newcommand{\araa}{ARA\&A}   \newcommand{\aap}{A\&A}
\newcommand{\aj}{AJ}         \newcommand{\apj}{ApJ}
      \newcommand{\apjs}{ApJS}
\newcommand{\mnras}{MNRAS}   
     \newcommand{\pasp}{PASP}
\newcommand{\procspie}{Proc.\ SPIE}  \newcommand{\aaps}{A\&AS}

\newcommand{\visclass}{\textsc{vis\_class}}

\newcommand{\newtext}[1]{{#1}}  

\title[GAMA: The input catalogue and star-galaxy separation]
{Galaxy And Mass Assembly (GAMA): 
The input catalogue and star-galaxy separation}

\author[I.~K.~Baldry et al.]
{I.~K.~Baldry$^{1}$, 
A.~S.~G.~Robotham$^{2}$, 
D.~T.~Hill$^{2}$, 
S.~P.~Driver$^{2}$, 
J.~Liske$^{3}$,
P.~Norberg$^{4}$,\newauthor
S.~P.~Bamford$^{5}$,
A.~M.~Hopkins$^{6}$,
J.~Loveday$^{7}$,
J.~A.~Peacock$^{4}$,
E.~Cameron$^{2,8}$,\newauthor
S.~M.~Croom$^{9}$,
N.~J.~G.~Cross$^{4}$, 
I.~F.~Doyle$^{10}$,
S.~Dye$^{11}$, 
C.~S.~Frenk$^{12}$,
D.~H.~Jones$^{6}$,\newauthor
E.~van Kampen$^{3}$,
L.~S.~Kelvin$^{2}$,
R.~C.~Nichol$^{10}$,
H.~R.~Parkinson$^{4}$,
C.~C.~Popescu$^{13}$,\newauthor
M.~Prescott$^{1}$, 
R.~G.~Sharp$^{6}$,
W.~J.~Sutherland$^{14}$, 
D.~Thomas$^{10}$, 
R.~J.~Tuffs$^{15}$\\ 
$^{1}$ Astrophysics Research Institute, Liverpool John Moores University, 
  Twelve Quays House, Egerton Wharf, Birkenhead CH41~1LD\\
$^{2}$ Scottish Universities' Physics Alliance (SUPA), School of Physics and Astronomy, 
  University of St Andrews, North Haugh, St Andrews, Fife, KY16 9SS\\
$^{3}$ European Southern Observatory, Karl-Schwarzschild-Str.~2, 85748 Garching, Germany\\
$^{4}$ SUPA, Institute for Astronomy, University of Edinburgh, Royal Observatory, Blackford Hill, 
  Edinburgh EH9 3HJ\\
$^{5}$ Centre for Astronomy and Particle Theory, University of Nottingham, University Park, 
  Nottingham NG7 2RD\\
$^{6}$ Anglo Australian Observatory, PO Box 296, Epping NSW 1710, Australia\\
$^{7}$ Astronomy Centre, University of Sussex, Falmer, Brighton, BN1 9QH\\
$^{8}$ Department of Physics, Swiss Federal Institute of Technology (ETH-Z\"urich), 
  CH-8093 Z\"urich, Switzerland\\
$^{9}$ Sydney Institute for Astronomy, School of Physics, University of Sydney, NSW 2006, Australia\\
$^{10}$ Institute of Cosmology and Gravitation (ICG), Dennis Sciama Building, Burnaby Road, 
  University of Portsmouth, PO1 3FX\\
$^{11}$ School of Physics \& Astronomy, Queens Buildings, The Parade, Cardiff University, CF24 3AA\\
$^{12}$ Institute for Computational Cosmology, Department of Physics, Durham University, South Road, 
  Durham DH1 3LE\\
$^{13}$ Jeremiah Horrocks Institute, University of Central Lancashire, Preston PR1 2HE\\
$^{14}$ Astronomy Unit, Queen Mary University London, Mile End Rd, London E1 4NS\\
$^{15}$ Max Planck Institute for Nuclear Physics (MPIK), Saupfercheckweg 1, D-69117 Heidelberg, Germany
}

\begin{document}

\date{MNRAS, accepted 2010 January 04. Received 2009 December 17; 
  in original form 2009 September 24}

\pagerange{\pageref{firstpage}--\pageref{lastpage}} \pubyear{2009}

\maketitle

\label{firstpage}

\begin{abstract}
We describe the spectroscopic target selection for the Galaxy And Mass
Assembly (GAMA) survey. The input catalogue is drawn from the Sloan Digital
Sky Survey (SDSS) and UKIRT Infrared Deep Sky Survey (UKIDSS).  The initial 
aim is to
measure redshifts for galaxies in three $4\times12$ degree regions at 9\,h,
12\,h and 14.5\,h, on the celestial equator, with magnitude selections
$r<19.4$, $z<18.2$ and $K_{\rm AB}<17.6$ over all three regions, and $r<19.8$
in the 12-h region. The target density is $1080\,\persqdeg$ in the 12-h region
and $720\,\persqdeg$ in the other regions.  The average GAMA target density
and area are compared with completed and ongoing galaxy redshift surveys.  The
GAMA survey implements a highly complete star-galaxy separation that jointly
uses an intensity-profile separator ($\dsg=r_{\rm psf}-r_{\rm model}$ as per
the SDSS) and a colour separator.  The colour separator is defined as
$\dsgjk=J-K-f(g-i)$, where $f(g-i)$ is a quadratic fit to the $J-K$ colour of
the stellar locus over the range $0.3<g-i<2.3$.  All galaxy populations
investigated are well separated with $\dsgjk>0.2$.  From two years out of a
three-year AAOmega program on the Anglo-Australian Telescope, we have obtained
79\,599 unique galaxy redshifts.  Previously known redshifts in the GAMA
region bring the total up to 98\,497.  The median galaxy redshift is 0.2 with
99\% at $z<0.5$. We present some of the global statistical properties of the
survey, including \newtext{$K$-band galaxy counts}, colour-redshift relations 
and preliminary $n(z)$.
\end{abstract}


\section{Introduction}
\label{sec:intro}

Galaxy redshift surveys provide a fundamental resource for studies of galaxy
evolution. The redshift of a galaxy can be used to obtain a distance assuming
a set of cosmological parameters, modulo peculiar velocities, and a
well-defined selection function enables the comoving number density of
galaxies to be estimated as a function of various properties, e.g., galaxy
luminosity functions \citep{Schechter76,BST88,norberg02,blanton03ld}.  In
addition, using the combined sky distribution and distance information, the
clustering properties of galaxies can be determined
\citep{DGH78,dLGH88,norberg02clus,zehavi05} and the velocity dispersion of
galaxies in groups and clusters can be used to infer dark-matter halo masses
\citep{Zwicky37,HG82,MFW93,carlberg96,eke04,berlind06}.

\begin{figure}
\centerline{
\includegraphics[width=\singlecolsize\textwidth]{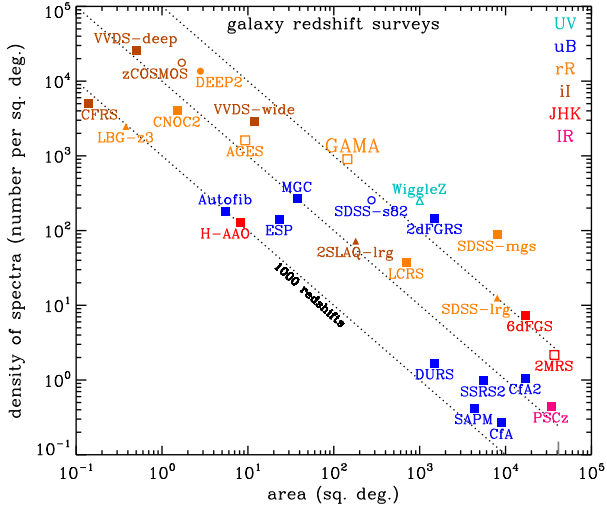}
}
\caption{Comparison between field galaxy surveys with spectroscopic
  redshifts: {\it squares} represent predominantly magnitude-limited
  surveys; {\it circles} represent surveys involving colour cuts for
  photometric redshift selection; while {\it triangles} represent
  highly targeted surveys. The colours represent different principal
  wavelength selections as in the legend. Filled symbols represent
  completed surveys. See Table~\ref{tab:z-survey-list} for survey names
  and references.}
\label{fig:compare-grs}
\end{figure}

The target selection algorithm and area covered by a redshift survey relate to
the redshift range and volume surveyed.  The industry of these surveys started
in the 1980's with surveys of $\sim2500$ galaxies over large sky areas
\citep{davis82,saunders90} and a deeper survey of 330 galaxies over
$70\,\sqdeg$ \citep{peterson86}. It expanded and diversified in the 1990's
with surveys such as the wide-but-shallow CfA2 redshift survey, Las Campanas
Redshift Survey, ESO Slice Project, and the deep-but-narrow Canada-France
Redshift Survey.  Figure~\ref{fig:compare-grs} shows the surface density of
galaxy spectra versus area for these and other surveys, and
Table~\ref{tab:z-survey-list} gives selections and references. The target
density is a wavelength-independent metric for depth, at least for
high-completeness magnitude-limited surveys. The advent of multi-object
spectrographs such as the Two-Degree Field (2dF; \citealt{lewis02df}) and
Sloan Digital Sky Survey (SDSS; \citealt{york00}) telescope have enabled
redshift surveys of $>10^5$ galaxies: the 2dF Galaxy Redshift Survey (2dFGRS)
and SDSS Main Galaxy Sample (MGS).

\begin{table*}
\caption{List of field galaxy redshift surveys. The surveys shown in
  Fig.~\ref{fig:compare-grs} are listed in order of increasing
  area. They are mostly magnitude limited galaxy samples except for
  some with colour selection (CS). The information was obtained from
  the references and the survey websites.}
\label{tab:z-survey-list}
\begin{tabular}{lllcl} \hline
abbrev.  & survey name  &  selection(s) & area/$\sqdeg$ & reference \\ \hline
CFRS     & Canada-France Redshift Survey & $I_{\rm AB}<22.5$ & $0.14$ & \citealt{lilly95} \\
LBG-z3   & Lyman Break Galaxies at $z\sim3$ Survey & $R_{AB}<25.5$ with CS$^a$ & $0.38$ & \citealt{steidel03} \\
VVDS-deep& VIMOS VLT Deep Survey deep sample & $I_{\rm AB}<24.0$ & $0.5$ & \citealt{lefevre05} \\
CNOC2    & Canadian Network for Obs.\ Cosmology 2 \rlap{...} & $R<21.5$ & $1.5$ & \citealt{yee00} \\
zCOSMOS  & Redshifts for the Cosmic Evolution Survey & $I_{\rm AB}<22.5$, $I_{\rm AB}\la24$ with CS$^b$  & $1.7$ & \citealt{lilly07} \\
DEEP2    & Deep Evolutionary Exploratory Probe 2 \rlap{...} & $R_{\rm AB}<24.1$ with CS$^c$ & $2.8$ & \citealt{davis03} \\
Autofib  & Autofib Redshift Survey & $b_{J}<22.0$ & $5.5$ & \citealt{ellis96} \\
H-AAO    & Hawaii+AAO K-band Redshift Survey & $K<15.0$ & $8.2$ & \citealt{huang03} \\
AGES     & AGN and Galaxy Evolution Survey & incl.\ $R<20.0$, $B_{\rm W}<20.5$ & $9.3$ & \citealt{watson09} \\ 
VVDS-wide& VIMOS VLT Deep Survey wide sample & $I_{\rm AB}<22.5$ & $12.0$ & \citealt{garilli08}\\
ESP      & ESO Slice Project & $b_{J}<19.4$ & $23.3$ & \citealt{vettolani97}\\
MGC      & Millennium Galaxy Catalogue & $B<20.0$ & $37.5$ & \citealt{liske03}\\
{\bf GAMA}&Galaxy And Mass Assembly Survey & $r<19.8$, $z<18.2$, $K_{\rm AB}<17.6$ & $144$ & --- this paper --- \\
2SLAQ-lrg& 2SLAQ Luminous Red Galaxy Survey & $i<19.8$ with CS$^d$ & $180$ & \citealt{cannon06}\\
SDSS-s82 & SDSS Stripe 82 surveys & incl.\ $u\la20$, $r<19.5$ with CS$^e$ & $275$ & \citealt{sdssDR4}\\
LCRS     & Las Campanas Redshift Survey & $R<17.5$ & $700$ & \citealt{shectman96}\\
WiggleZ  & WiggleZ Dark Energy Survey & ${\rm NUV}<22.8$ with CS$^f$ & $1000$ & \citealt{drinkwater09}\\
2dFGRS   & 2dF Galaxy Redshift Survey & $b_{J}<19.4$ & $1500$ & \citealt{colless01}\\
DURS     & Durham-UKST Redshift Survey & $b_{J}<17.0$ & $1500$ & \citealt{ratcliffe96}\\
SAPM     & Stromlo-APM Redshift Survey & $b_{J}<17.1$ (1 in 20 sampling) & $4300$ & \citealt{loveday92}\\
SSRS2    & Southern Sky Redshift Survey 2 & $B < 15.5$ & $5500$ & \citealt{dacosta98}\\
SDSS-mgs & SDSS Main Galaxy Sample & $r<17.8$ & $8000$ & \citealt{strauss02}\\
SDSS-lrg & SDSS Luminous Red Galaxy Survey & $r<19.5$ with CS$^g$ & $8000$ & \citealt{eisenstein01} \\
6dFGS    & 6dF Galaxy Survey & $K<12.7$, $b_{J},r_F,J,H$ limits & $17000$ & \citealt{jones09} \\
CfA2     & Center for Astrophysics 2 Redshift Survey & $B<15.5$ & $17000$ & \citealt{falco99}$^h$ \\
PSCz     & IRAS Point Source Catalog Redshift Survey & $60\mu m_{\rm AB}<9.5$ & 34000 & \citealt{saunders00} \\
2MRS     & 2MASS Redshift Survey & $K<12.2$ & 37000 & \citealt{erdogdu06} \\ \hline
\end{tabular}
\begin{flushleft}
Notes: 
$^a$CS by $U$-band `dropouts' for photometric redshifts ($z_{\rm ph}$) $\sim2.5$--3.5; 
$^b$CS for $z_{\rm ph}\sim1.4$--3.0, deeper limit over $1\,\sqdeg$; 
$^c$CS for $z_{\rm ph}\ga0.7$; 
$^d$CS for $z_{\rm ph}\sim0.45$--0.8; 
$^e$CS for $z_{\rm ph}\la0.15$; 
$^f$CS by ${\rm FUV-NUV}>1.5$ (GALEX bands) and $20.5<r<22.5$ for $z_{\rm ph}\sim0.5$--1.0; 
$^g$CS for $z_{\rm ph}\sim0.2$--0.5; 
$^h$Reference is for the Updated Zwicky Catalog that includes CfA2 redshifts. 
\end{flushleft}
\end{table*}

The Galaxy And Mass Assembly (GAMA) project has at its core a galaxy redshift
survey using the upgraded 2dF instrument AAOmega on the Anglo-Australian
Telescope (AAT). GAMA will eventually incorporate a range of new surveys from UV,
visible, IR and radio wavelengths \citep{driver09}.  The redshift survey uses
for its input catalogue data from the Sloan Digital Sky Survey and United
Kingdom InfraRed Telescope (UKIRT).  The primary goals of the redshift survey
are measurement of the halo mass function \citep{eke06}, galaxy stellar mass
function \citep{cole01}, and the merger rates of galaxies \citep{depropris05}:
for systems of the lowest possible masses (at low redshift $z<0.05$), and for
their evolution out to $z\sim0.5$.  In terms of depth and area of
magnitude-limited surveys (squares in Fig.~\ref{fig:compare-grs}), GAMA
bridges the gap between the wide but shallower surveys like 2dFGRS and SDSS
MGS, and the deep but narrower surveys such as those using the VIsible
MultiObject Spectrograph (VIMOS) on the Very Large Telescope (VLT).

The outline of the paper is as follows.  The imaging data, magnitude
measurements and initial catalogues are described in \S~\ref{sec:imaging}.
The target selection is described in \S~\ref{sec:target-select}: star-galaxy
separation, magnitude limits, and other quality checks.  The pre-existing and
GAMA spectroscopic data sets are outlined in \S~\ref{sec:spec}. An analysis of
results as pertaining to the star-galaxy separation and other selection
criteria is presented in \S~\ref{sec:results}. In other survey papers, the
scientific and multi-wavelength database aims are described in
\citet{driver09}, and the tiling strategy is described in \citet{robotham09}.

Magnitudes are corrected for Milky-Way extinction using the dust maps
of \citet{SFD98} except for fibre magnitudes. Extinction in the bands
$u,g,i,z,J,K$ are obtained from SDSS $r$-band extinction using fixed
ratios (1.873864, 1.378771, 0.758270, 0.537623, 0.323,
0.131).\footnote{The SDSS extinction ratios are given in table~22 of
  \citet{stoughton02}.  The ratios for $J$- and $K$-band extinctions
  were obtained from UKIRT WFCAM science archive \citep{hambly08} data
  matched to SDSS $r$-band extinction.}  The UKIRT magnitudes are
converted to the AB system using $J_{\rm AB} = J + 0.94$ and $K_{\rm
  AB} = K + 1.90$ \citep{hewett06}.  The contours used to represent
bivariate distributions
(Figs.~\ref{fig:sg-sep-histo}--\ref{fig:test-auto-jk},
\ref{fig:color-bias}--\ref{fig:size-z}, \ref{fig:obs-color-z}) are
logarithmically spaced in number density, with four levels per factor
of ten.

\section{Imaging}
\label{sec:imaging}

\subsection{Sloan Digital Sky Survey and GAMA regions}
\label{sec:sdss}

The SDSS project \citep{york00,stoughton02} has used a dedicated 2.5-m
telescope to image $\sim10^4\,\sqdeg$ and to obtain spectra of $\sim10^6$
objects \citep{sdssDR6}.  The imaging was obtained through five broadband
filters, {\it ugriz} with effective wavelengths of 355, 470, 620, 750 and
895\,nm, using a mosaic CCD camera consisting of 5 rows and 6 columns
\citep{gunn98}.  Observations with a 0.5-m photometric telescope
\citep{hogg01} are used to calibrate the 2.5-m telescope images using the
$u'g'r'i'z'$ standard star system \citep{fukugita96,smith02}.  The GAMA survey
targets were selected using DR6 imaging.\footnote{We are aware that a new
  photometric calibration was implemented for the DR7 release
  \citep{padmanabhan08,sdssDR7}. However, the magnitude changes are typically
  less than 0.02\,mag and therefore, for consistency, we have not used the DR7
  magnitudes because we started spectroscopic observations prior to this
  release.}

\begin{figure*}
\includegraphics[width=\middlecolsize\textwidth]{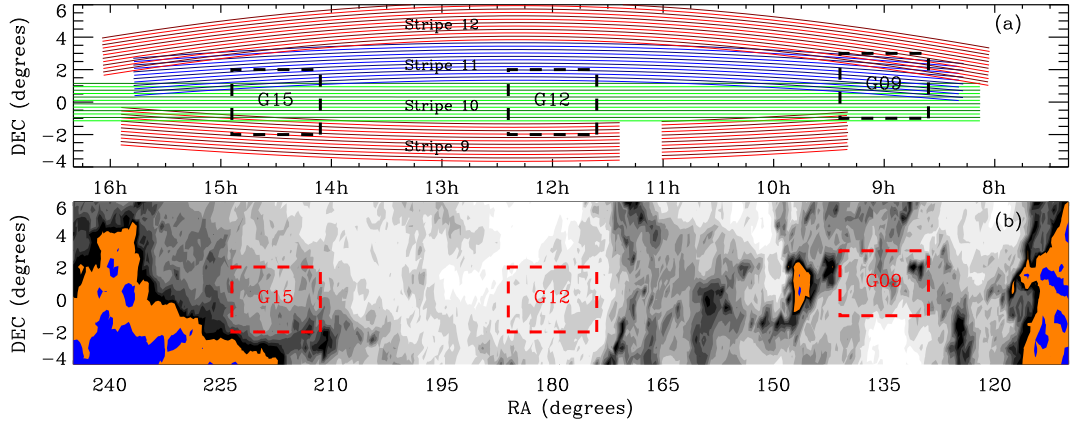}
\caption{{\bf (a):} Scan-line positions for SDSS Stripes 9--12.  The GAMA
  regions are outlined using dashed lines.  The twelve scan-lines for each
  stripe are the result of interleaving North and South strips each with six
  camera columns. {\bf (b):} GAMA regions in relation to the dust map of
  \citet{SFD98}. The colours represent SDSS $r$-band extinction in magnitude
  ranges: $<0.06$ white; 0.06-0.20 grey scale; 0.20-0.25 black; 0.25--0.5
  orange; and $>0.5$ blue.}
\label{fig:sdss-stripes}
\end{figure*}

The imaging was obtained by drift scanning along a strip defined in an SDSS
coordinate system. Two strips, designated N and S, are interleaved to fill in
the gaps between the camera columns and are combined to make one stripe. The
choice for the GAMA survey consisted of the Southern-most stripes (${\rm DEC}
< 3\degr$)
for good access from Southern observatories. The contiguous SDSS coverage of
Stripes 9--12 was chosen to allow GAMA regions that are four-degrees wide: an
estimated requirement for group finding and measurement of the halo mass
function at $z<0.1$ \citep{driver09}. Figure~\ref{fig:sdss-stripes} shows
these regions in relation to the SDSS stripes and Milky-Way extinction. They
each cover $4\times12$ degrees and are centred on 9\,h, 12\,h and 14.5\,h.
The RA and DEC ranges are given in Table~\ref{tab:gama-regions}.

\begin{table}
\caption{The GAMA regions defined in J2000 coordinates}
\label{tab:gama-regions}
\begin{tabular}{lcc} \hline
{\bf G09}   & $129\degr.0<{\rm RA}<141\degr.0$ & $-1\degr.0<{\rm DEC}<3\degr.0$ \\
{\bf G12}   & $174\degr.0<{\rm RA}<186\degr.0$ & $-2\degr.0<{\rm DEC}<2\degr.0$ \\ 
{\bf G15}   & $211\degr.5<{\rm RA}<223\degr.5$ & $-2\degr.0<{\rm DEC}<2\degr.0$ \\ \hline
\end{tabular}
\end{table}

The SDSS produces various magnitude measurements
\citep{stoughton02}. These include:
\begin{itemize}
\item Petrosian magnitudes measured using a circular aperture that
  is twice the Petrosian radius.  The radius is determined using the
  surface brightness profile of the object in the $r$-band.
\item Model magnitudes determined from the best fit of an
  exponential or de Vaucouleurs profile. The shape parameters
  (major-minor axes ratio, position angle, scale radius) are
  determined from the $r$-band image, while only the amplitude is
  fitted in the other bands.
\item PSF magnitudes determined from a fit using the point spread
  function in each band.
\item Fibre magnitudes measured using a circular aperture that is
  $3''$ in diameter. For these magnitudes, no attempt is made to
  deblend overlapping objects. Their purpose is to provide an estimate
  of signal in the spectrographs.
\end{itemize}
These magnitude types are all used in our selection for various
reasons.  Note we make no adjustment from the SDSS $3''$ fibre
magnitudes to AAOmega $2''$ apertures. An average correction is
0.35\,mag, with the 95\% range from 0.15--0.6\,mag (for galaxies with
$18<r<20$).

The SDSS pipeline \textsc{photo} also gives a number of flags for each
measured source (table~9 of \citealt{stoughton02}).  The most important for
target selection is \textsc{satur}, which is set if any pixel in a source or
its `parent' is saturated. This can be used to effectively exclude deblends of
bright stars. We also consider the \textsc{parentid} of sources, which can be
used to group together objects that may be significantly overlapping. This is
used in the visual classification process (\S~\ref{sec:vis-class}) to identify
deblended parts of galaxies.

The initial input catalogue was selected from the \textsc{dr6.PhotoObj}
table with, in addition to magnitude limits and area restrictions,
the following criteria (in SQL):{\small
\begin{verbatim}
(mode = 1)  or  
(mode = 2 and ra < 139.939 and dec < -0.5 and 
  (status & dbo.fphotostatus('OK_SCANLINE')) > 0)
\end{verbatim}
}\noindent The \textsc{mode} column is set to 1 for primary objects and 2 for
secondary objects, which are in areas where stripes and/or scan-lines overlap.
However, Stripe~9 is mostly incomplete for G09 and thus secondary objects need
to be selected from some Stripe~10 scan-lines in this region because the code
assumes Stripe~9 is complete when determining the \textsc{mode} values [see
  Fig.~\ref{fig:sdss-stripes}(a), consider the extension of Stripe~9 to 8\,h].
The RA and DEC limits above select the appropriate part of Stripe~10, and the
\textsc{ok\_scanline} flag ensures that selected objects are not in the
overlap edge areas of the scan-lines.

While data from Stripes 9--12 were used for GAMA target selection, data from
from Stripe~82 were used for early testing of our star-galaxy separation
method.  This was because of the available UKIRT $J$ and $K$ band coverage at
the time and because of significant additional SDSS redshifts beyond the main
SDSS surveys. The additional targets included selections for both resolved and
unresolved sources \citep{sdssDR4}.

\subsection{UKIRT Infrared Deep Sky Survey}
\label{sec:ukirt}

The UKIRT Infrared Deep Sky Survey (UKIDSS; \citealt{dye06,lawrence07}) is a
project using the Wide-Field Camera (WFCAM; \citealt{casali07}) on the 3.8-m
UKIRT. The WFCAM instrument consists of four $2k\times2k$ HgCdTe detectors in
a two-by-two pattern. Each detector covers $13.7'\times13.7'$ and is separated
from neighbouring detectors by $12.9'$ (94\% of each detector's active
length). Thus, four observations can be interleaved to form a contiguous
$0.9\degr\times0.9\degr$ tile.  The available filters are {\it ZYJHK} with
effective wavelengths of 0.88, 1.03, 1.25, 1.63 and $2.20\,\mu m$
\citep{hewett06}.  The UKIDSS consists of a number of different sub-surveys,
including the Large Area Survey (LAS) obtaining imaging in {\it YJHK} over
$>2000\,\sqdeg$ within the SDSS main survey regions.

There is a dedicated pipeline for reducing and a system for archiving the
UKIDSS data \citep{hambly08}.  However, we did not use the fully reduced data
product catalogues for the GAMA regions when we incorporated UKIDSS LAS data
into our selection criteria. This was partly because of known problems with
the deblending algorithm, and also our desire to have control over aperture
matched photometry.  Reduced LAS images, the detector frames, were obtained
from the archive.  These were scaled to a common background and gain, and
$YJHK$ mosaics were produced using the AstrOmatic SWarp program
\citep{bertin02}. \newtext{A systematic study of the calibration errors in
  \citet{hodgkin09} finds that the photometry is accurate to better than 0.02
  magnitudes rms when tested against the Two Micron All Sky Survey 
  (2MASS; \citealt{skrutskie06}) for J, H and K bands, making a global 
  recalibration unnecessary.}

\begin{figure}
\centerline{
\includegraphics[width=\singlecolsize\textwidth]{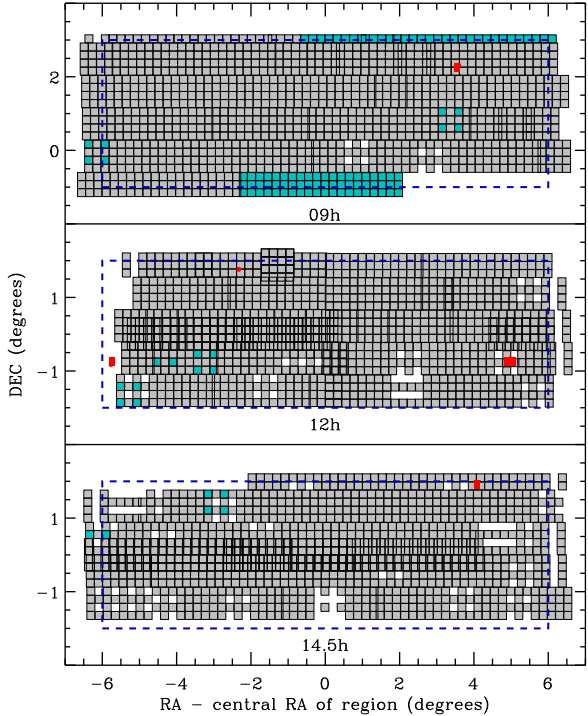}
}
\caption{UKIDSS $J$ and $K$-band coverage for AAT observations in 2009.  The
  squares represent the WFCAM frames: filled grey if $J$ and $K$- bands were
  available, and cyan if only the $K$-band was available.  \newtext{The
    $K$-band areal coverage of G09, G12 and G15 is 96.5\%, 93.5\% and 88.2\%,
    respectively.}  The red areas show the largest areas missing from the SDSS
  coverage because of masking around the brightest stars ($3.5<V<4.5$; HR3665,
  HR4471, HR4540, HR4689, HR5511) and related `timed out' frames.}
\label{fig:ukidss-coverage}
\end{figure}

Each GAMA region has pixel aligned 20\,GB mosaics for each band, alleviating
problems due to multiple edge extractions and allowing us to use matched
aperture photometry. \newtext{The final mosaics have a $0.4''$/px scale, use
  median co-addition in overlap regions and interpolate the resampled pixels
  using Lanczos resampling level 3. These latter setting is as suggested in
  the SWarp manual.  The use of matched aperture photometry is important for
  improving the quality of the galaxy colours, and our star-galaxy
  separation.}  \textsc{SExtractor} \citep{BA96} was run in dual mode on the
$J$ and $K$ images, with the source positions and sizes defined in the $K$
band, using default parameters. This catalogue was then matched to an initial
SDSS catalogue (for GAMA) within a $2''$ tolerance using \textsc{stilts}
\citep{Taylor05}, with the nearest match chosen when there were multiple
matches.  Figure~\ref{fig:ukidss-coverage} shows the $J$ and $K$-band LAS
coverage used for target selection prior to AAT observations in 2009.  While
the UKIDSS coverage will be completed and may be used for
future targeting, any analysis considering completeness as a function of
position will need to take account of the UKIDSS coverage prior to the 2009
observations.

The output from \textsc{SExtractor} gives a number of flux measurements.  Here
we generally use the standard \textsc{auto} magnitude, based on an elliptical
aperture defined using \citeauthor{Kron80}'s (\citeyear{Kron80}) algorithm.
These \textsc{auto} magnitudes are used for $K$-band selection and for $J-K$
colours as part of the star-galaxy separation criteria.  For early tests of
our star-galaxy separation using Stripe~82 data (\S~\ref{sec:star-gal-sep}),
we used the available UKIDSS pipeline \textsc{apermag3} measurements, which
are determined using $2.0''$ circular apertures.

\newtext{The fidelity of additional UKIDSS targets is a key concern.  The
  imaging used in this work has already passed the basic quality assessments
  discussed in \citet{dye06} and \citet{warren07}. This involves removing
  completely corrupted data and images affected by moon ghosting and other
  serious visual artefacts. To further ensure the quality of additional
  near-IR selected objects two precautions were taken.  Firstly all targets
  had to possess a SDSS counterpart (see above), and secondly selected UKIDSS
  targets were visually inspected by a small group within the
  GAMA team (\S~\ref{sec:vis-class}).  This level of care goes a great way to
  mitigating against any data quality issues in the original UKIDSS data.  The
  fraction of fibres placed on spurious artefacts should be small since they
  would have to be present in SDSS and UKIDSS imaging at the same position in
  the sky.}

\begin{figure}
\centerline{
\includegraphics[width=\smallcolsize\textwidth]{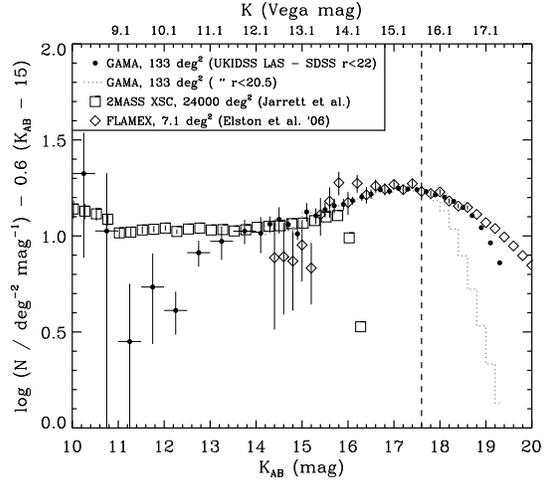}
}
\caption{\newtext{$K$-band galaxy counts for the UKIDSS GAMA regions, 2MASS
    Extended Source Catalog and FLAMINGOS Extragalactic Survey.  The $y$-axis
    shows the logarithmic counts with the slope for a non-evolving Euclidean
    universe subtracted.  The dotted line shows the drop in counts at the
    faint end when the SDSS match is restricted to $r_{\rm model}<20.5$. The
    vertical dashed line is the limit used for the $K$-band selection
    (\S~\ref{sec:mag-limits}). The GAMA errors were determined from standard
    errors using 6 different areas (2 per GAMA region); while the errors for
    the other surveys were obtained from tables provided by T.~Jarrett and
    A.~Gonzalez, respectively.}}
\label{fig:K-counts}
\end{figure}

\newtext{Figure~\ref{fig:K-counts} shows the $K$-band galaxy counts using the
  derived \textsc{auto} magnitudes in the GAMA regions both from a match to an
  SDSS $r<22$ catalogue and an $r<20.5$ catalogue.  There is excellent
  agreement between the UKIDSS LAS and 2MASS counts
  \citep{jarrett00,jarrett10} from about 12.5 to 15.8 AB mag. At brighter
  magnitudes, the discrepancy is not of concern for targeting because these
  bright galaxies will have redshifts anyway; the discrepancy could be caused
  by cosmic variance and/or the use of the $2''$ tolerance. At fainter
  magnitudes, the 2MASS incompleteness is evident, while the UKIDSS counts are
  in good agreement with the FLAMINGOS Extragalactic Survey (FLAMEX;
  \citealt{elston06}) counts to $K_{\rm AB}\la18.8$. The agreement in the
  galaxy counts between these surveys demonstrates consistency in the derived
  magnitudes and star-galaxy separation, which for GAMA is described in the
  following section.}

\section{Target selection}
\label{sec:target-select}

\subsection{Star-galaxy separation}
\label{sec:star-gal-sep}

\begin{figure}
\centerline{
\includegraphics[width=\smallcolsize\textwidth]{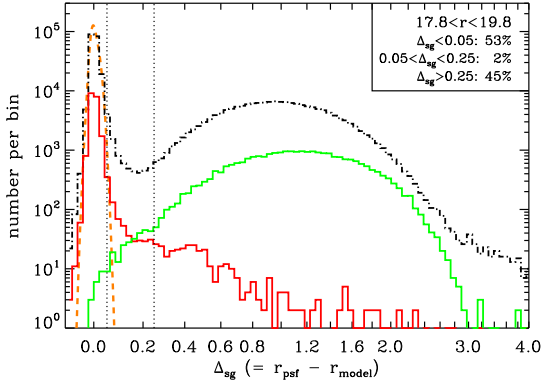}
}
\caption{Histogram of $\dsg$ for Stripe~82 data. The $x$-axis stretch is
  linear in $\ln(1+\dsg)$ (bin size is 0.02).  The dash-dotted histogram
  represents all objects, while the red histogram represents objects with
  confirmed stellar redshifts ($-0.002<z<0.002$), and the green histogram
  represents extra-galactic sources with $0.002<z<0.35$.  Note the $y$-axis is
  on a logarithmic scale and the orange dashed curve shows a double Gaussian
  fit to the stellar peak (${\rm FWHM} \sim 0.025$). The vertical dotted lines
  show the range for marginally-resolved sources ($0.05 < \dsg < 0.25$).}
\label{fig:sg-sep-histo}
\end{figure}

Automatic separation of stars and galaxies from images has typically
been done using shape or intensity profile measurements (e.g.,
\citealt{macgillivray76,maddox90}).  The SDSS star-galaxy separation
parameter \citep{strauss02} is defined as
\begin{equation}
  \dsg = r_{\rm psf} - r_{\rm model} 
\label{eqn:sdss-star-gal-sep}
\end{equation}
where $r_{\rm psf}$ and $r_{\rm model}$ are the $r$-band PSF and model
magnitudes.  The value deviates from zero when the de Vaucouleurs or
exponential profile fit accounts for more flux than only using a PSF fit,
i.e., a significant deviation from zero indicates that the intensity profile
is not well matched to the PSF.  Figure~\ref{fig:sg-sep-histo} shows a
histogram in this parameter for objects with $17.8<r_{\rm petro}<19.8$ that
are not deblended from a saturated object. Also shown are objects with
confirmed stellar redshifts and galaxies with $0.002<z<0.35$ (from Stripe~82).
The cut $\dsg > 0.24$ was the constraint used for star-galaxy separation in
the SDSS MGS.\footnote{$\dsg > 0.3$ is the MGS criteria quoted in
  \citet{strauss02} but the limit was later reduced to 0.24 following the
  change in the model magnitude code at DR2 \citep{sdssDR2}.}  With this
selection, some galaxies that are compact will be missed, particularly as we
target fainter than $r=17.8$.

Our first cut is to select objects with $\dsg > 0.05$ (nominally marginally or
well resolved).  This removes the Gaussian core of objects that are
unresolved, which are almost all stars and quasars.  However, this cut is
still too inclusive of stars for targeting efficiency so further cuts need to
be applied.  In particular, the $0.05 < \dsg < 0.25$ region probably includes
many double-star systems as well as marginally-resolved galaxies. The latter
are selected using colour cuts based on our UKIDSS-SDSS matched
catalogue.\footnote{Note that even with UKIDSS-SDSS colour selection selecting
  objects with $\dsg < 0.05$ would result in a large stellar contamination to
  our galaxy sample. \newtext{The median PSF full-width half maximum for the
    SDSS $r$-band is $1.4''$ in the GAMA regions, with 95\% of `fields' having
    seeing better than $2.1''$. While the UKIDSS LAS $K$-band seeing is
    typically better than $\sim1''$, we have not yet modelled the PSF
    variation accurately and thus prefer to use the well-established SDSS
    profile separator rather than one based on UKIDSS (which in any case does
    not cover all the GAMA area).  Our UKIDSS-SDSS colour selection will
    mitigate against variation in the reliability of the SDSS profile
    separation.}}

\begin{figure}
\centerline{
\includegraphics[width=\singlecolsize\textwidth]{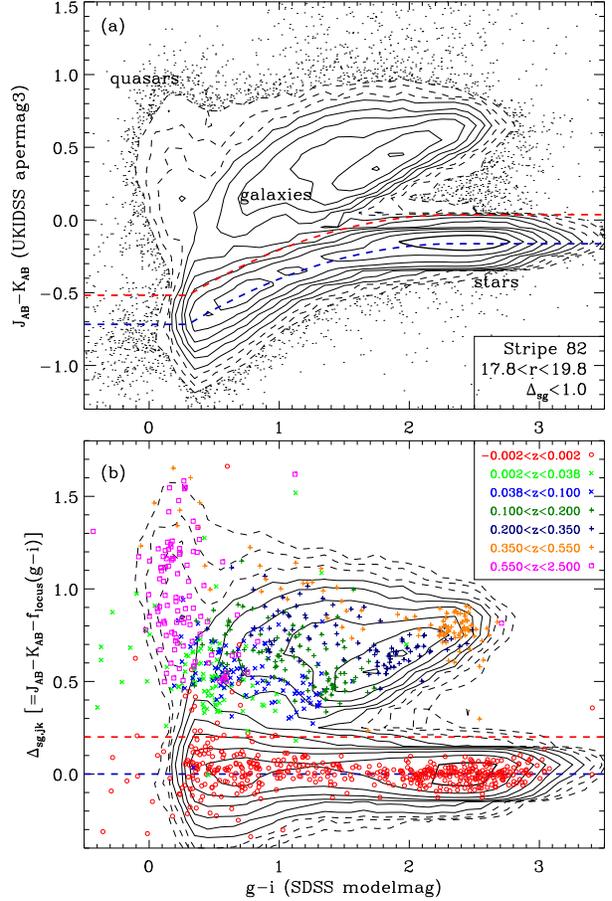}
}
\caption{{\bf (a):} Star-galaxy separation in colour-colour space.  The blue
  dashed line represents a fit to the stellar locus over the range
  $0.3<g-i<2.3$ and constant $J-K$ either side of the fitted range
  (Eq.~\ref{eqn:f_locus}); while the red dashed line is $+0.2$ in $J-K$ from
  this fit.  {\bf (b):} $J-K$ star-galaxy separation parameter versus $g-i$
  for populations over different redshift ranges.  From objects with measured
  redshifts in Stripe~82, 500 stars and 100 in each extra-galactic redshift
  range were selected at random.  The red line shows the $\dsgjk$ cut.}
\label{fig:ukidss-sg-sep}
\end{figure}

A UKIDSS-SDSS star-galaxy separation was determined using data from Stripe~82.
Figure~\ref{fig:ukidss-sg-sep}(a) shows a plot of $(J-K)_{\rm apermag3}$
versus $(g-i)_{\rm model}$ for objects with $\dsg < 1.0$ and $17.8 < r_{\rm
  petro} < 19.8$ (i.e.\ fainter than SDSS MGS within GAMA selection). A
colour-colour diagram using these bands was utilized by \citet{ivezic02} to
assess the success of SDSS star-galaxy separation \newtext{and similarly by
  \citet{elston06}, with $B_W-I$ instead of $g-i$, for the FLAMEX star-galaxy
  separation.}

From selected sources, we fit the stellar locus with a quadratic.  A
new star-galaxy separation parameter is defined as the $J-K$
separation from the locus, which is shown by the blue dashed line.
The parameter is given by
\begin{equation}
 \dsgjk = J_{\rm AB} - K_{\rm AB} - f_{\rm locus}(g-i)
\label{eqn:ukidss-sdss-star-gal-sep}
\end{equation}
where
\begin{equation}
 f_{\rm locus}(x) = 
\begin{array}{l} 
  -0.7172 \\ -0.89 + 0.615 x - 0.13 x^2 \\  -0.1632 \end{array}
\mbox{~for~} 
\begin{array}{l} 
  x < 0.3 \\ 0.3 < x < 2.3 \\ x > 2.3
\end{array}
\label{eqn:f_locus}
\end{equation}
Figure~\ref{fig:ukidss-sg-sep}(b) shows $\dsgjk$ versus $(g-i)_{\rm model}$,
with symbols representing samples that have measured redshifts. The cut
$\dsgjk > 0.20$ is used to select extra-galactic sources among the objects
with $0.05<\dsg<0.25$ (the success and completeness of this UKIDSS-SDSS
star-galaxy separation are presented later in \S~\ref{sec:results-sg-sep}).

\begin{figure}
\centerline{
\includegraphics[width=\smallcolsize\textwidth]{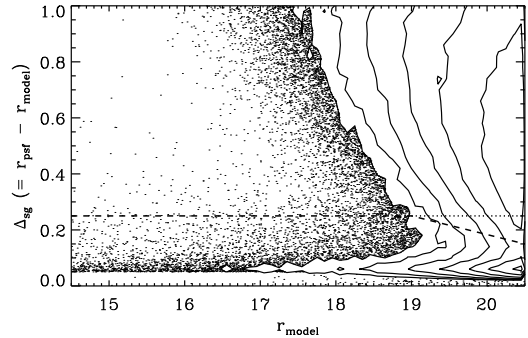}
}
\caption{Star-galaxy separation regions.  Above $\dsg=0.25$, objects are
  selected as a spectroscopic target regardless of $J-K$ colour (as per AAT
  observations in 2008 and SDSS MGS).  Below the dashed line,
  objects are selected if they satisfy the $J-K$ star-galaxy separation
  criteria ($\dsgjk>0.2$).  While above the dashed line but with $\dsg<0.25$
  (triangular region at $r>19$), objects are selected if $\dsgjk>0.2$ or there
  is no $J-K$ measurement.}
\label{fig:sg-sep}
\end{figure}

Not all objects have measured $J-K$. For these objects we lower the $\dsg$ cut
for fainter objects to $\dsg > f_{\rm sg,slope}(r_{\rm model})$ where
\begin{equation}
 f_{\rm sg,slope}(x) = 
\begin{array}{l} 
  0.25 \\ 0.25 - \frac{1}{15}(x - 19) \\  0.15 \end{array}
\mbox{~for~} 
\begin{array}{l} 
  x < 19.0 \\ 19.0 < x < 20.5 \\ x > 20.5
\end{array}
\end{equation}
Figure~\ref{fig:sg-sep} shows the distribution in $\dsg$ versus $r_{\rm
  model}$, with the cut shown by the red dashed line.  This is appropriate
because the sky density of objects that are galaxies compared to double stars,
in the marginally-resolved region, is increasing toward fainter magnitude
limits.

In summary, the overall star-galaxy separation is given by
\begin{equation}
\begin{array}{ll} 
  \dsg > 0.25                             & \mbox{\textsc{or}} \\
  \dsg > 0.05 \mbox{~~\textsc{and}~~} \dsgjk > 0.20 & \mbox{\textsc{or}} \\
  \dsg > f_{\rm sg,slope}(r_{\rm model}) \mbox{~\textsc{and}~ no $J-K$ measurement.}
\end{array}
\label{eqn:star-gal-summary}
\end{equation}
Only objects satisfying these criteria are targeted in the main survey.

\begin{figure}
\centerline{
\includegraphics[width=\singlecolsize\textwidth]{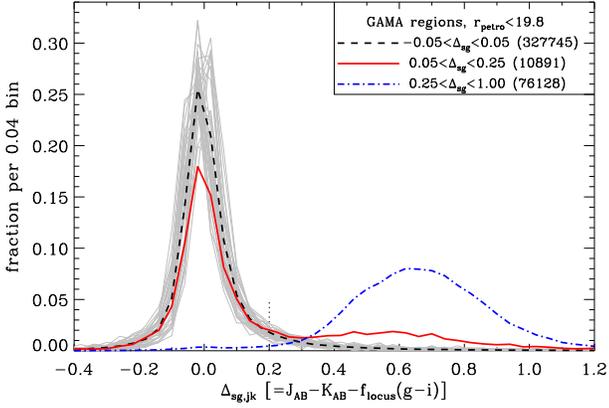}
}
\caption{\newtext{Histograms in $J-K$ star-galaxy separation parameter using
    \textsc{SExtractor} \textsc{auto} magnitudes.  The histograms represent
    UKIDSS-SDSS matched samples divided into unresolved (not selected for
    targeting), marginally resolved (selected if $\dsgjk>0.2$) and strongly
    resolved (selected regardless of $\dsgjk$) in SDSS $r$-band imaging. The
    thin grey line histograms represent unresolved samples in 36 randomly
    selected $1\degr \times 1\degr$ regions.}}
\label{fig:test-auto-jk}
\end{figure}

The GAMA UKIDSS selection was based on non-pipeline \textsc{SExtractor}
magnitudes. Thus, the final star-galaxy separation
(Eq.~\ref{eqn:ukidss-sdss-star-gal-sep}) was determined using \textsc{auto}
mags for $J-K$ and SDSS model mags for $g-i$ (data from Stripes 9--12).
\newtext{Figure~\ref{fig:test-auto-jk} shows histograms in $\dsgjk$ using
  these magnitudes. In order to test the position of the stellar locus,
  unresolved samples ($\dsg < 0.05$; not used for targeting) were selected in
  separate $1\degr \times 1\degr$ regions (grey lines in
  Fig.~\ref{fig:test-auto-jk}). The median value of $\dsgjk$ within each
  region varied from $-0.05$ to 0.04 with 90 per cent of the region values
  between $-0.03$ and 0.02. This demonstrates that the stellar locus fit
  applies to \textsc{auto} mags equally well.}

\subsection{Magnitude limits}
\label{sec:mag-limits}

The main scientific goal of GAMA that drives the choice of the minimum width
of the survey geometry, and the magnitude selection, is the measurement of the
halo mass function \citep{driver09}.  We chose $r$-band selection because it
is most directly correlated with spectral S/N obtained (the filter falls in
the middle range of the spectrograph). This ensures a high redshift success
rate for a given target density. The $r$-band limits were chosen to give an
average target density up to an order of magnitude higher than the SDSS MGS
($90\,\persqdeg$) and 2dFGRS ($140\,\persqdeg$). Given the limitations of
efficient observing over two or three lunations each year, three fields were
chosen covering 6 hours in RA. We compromised between area and depth by
choosing a limit of $r<19.4$ in G09 and G15 ($670\,\persqdeg$), and $r<19.8$
in G12 ($1070\,\persqdeg$). These were defined using Petrosian magnitudes,
following the strategy of the SDSS MGS.

\begin{figure}
\centerline{
\includegraphics[width=\smallcolsize\textwidth]{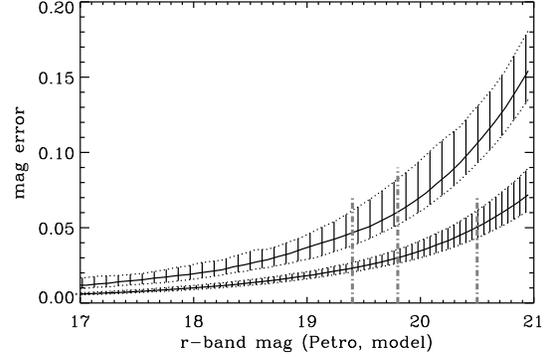}
}
\caption{Magnitude errors versus magnitude. The solid lines show the
  median errors obtained from the SDSS catalogue, with the regions
  representing the inter-quartile range (top for Petrosian, lower for
  model magnitudes). The vertical dash-dotted lines represent the
  $r$-band limits used in this paper (19.4 and 19.8 using Petrosian,
  and 20.5 using model magnitudes).}
\label{fig:mag-test}
\end{figure}

In consideration of measuring the stellar mass function, we included a near-IR
selection using SDSS $z$-band and UKIDSS $K$-band.  To ensure reliability and
reasonable redshift success rate, these were also constrained by an $r$-band
selection ($r_{\rm model}<20.5$).  The choice of SDSS model magnitudes rather
than Petrosian is a consequence of the noise statistics. For Petrosian
magnitudes, the noise is well behaved to $r\simeq20$ \citep{stoughton02},
while for fainter objects the model magnitudes are more
reliable. Figure~\ref{fig:mag-test} shows the pipeline-output magnitude errors
versus magnitude.  Also, the $K$-band selection was based on \textsc{auto}
magnitudes, and both \textsc{auto} and model magnitudes use elliptical
apertures.  The additional selections were a small sample to $z_{\rm
  model}<18.2$ and a sample to $K_{\rm AB,auto}<17.6$.

\begin{figure}
\centerline{
\includegraphics[width=\singlecolsize\textwidth]{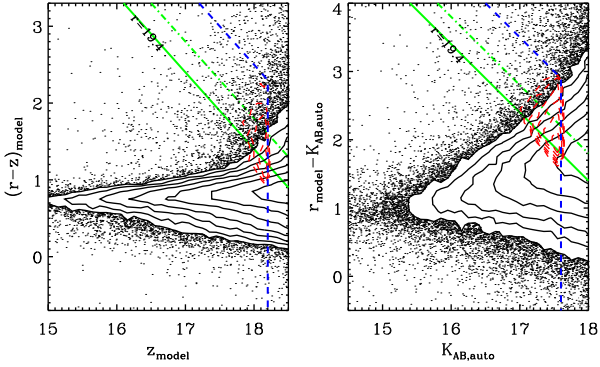}
}
\caption{Colour versus magnitude distribution for a near-IR sample.  The black
  contours and points represent potential galaxy targets.  The blue dashed
  lines show the limits imposed by our selection including the constraint
  $r_{\rm model}<20.5$.  The green lines show $r=19.4$ and 19.8 limits. The
  red contours represent most of the additional targets not selected by the
  $r_{\rm petro}$ limits. These contours extend below the solid green line
  because of differences between Petrosian and model magnitudes.}
\label{fig:color-bias}
\end{figure}

Within the GAMA regions, the main survey selections are given by:
\begin{equation}
\begin{array}{ll} 
r_{\rm petro}<19.4                            
  & \mbox{\textsc{or}} \\
r_{\rm petro}<19.8 \mbox{~~\textsc{and}~ in the G12 area} 
  & \mbox{\textsc{or}} \\
z_{\rm model}< 18.2 \mbox{~~\textsc{and}~~} r_{\rm model} < 20.5 
  & \mbox{\textsc{or}} \\
K_{\rm AB,auto}  < 17.6 \mbox{~~\textsc{and}~~} r_{\rm model} < 20.5  \: .
\end{array}
\label{eqn:mag-limits}
\end{equation}
Including the near-IR selections increases the G12 target density marginally
(to $1080\,\persqdeg$) while increasing the G09 and G15 target density to
$720\,\persqdeg$. Figure~\ref{fig:color-bias} shows the colour bias for the
near-IR selections.  The $z$-band selection is complete to $(r-z)_{\rm model}
< 2.3$ at the faint limit, while the $K$-band selection is complete to $r_{\rm
  model} - K_{\rm AB,auto} < 2.9$ at the faint limit. A $z_{\rm model}<18.2$
selection is formally missing 0.3\% of objects because of the $r_{\rm model}$
limit, while a $K_{\rm AB,auto}<17.6$ selection is formally missing about 1\%
of objects. This is after applying star-galaxy separation. However, only very
red objects are missed, which are more likely to be stars in spite of the
star-galaxy separation or have incorrectly measured colours caused by
mismatched apertures in the case of $r-K$ (the practical impact of these joint
limits is discussed later in \S~\ref{sec:z-distribution}).  \newtext{See also
  Fig.~\ref{fig:K-counts}, the $r<20.5$ limit only makes an obvious impact in
  the galaxy number counts at $K_{\rm AB,auto}>17.8$ that is above our
  selection limit.}

\subsection{Masking}
\label{sec:masking}

In order to avoid targeting galaxies with bad photometry because they are near
bright stars or satellite trails, an explicit mask was constructed.  The
bright-stars mask was based on stars down to $V<12$ in the Tycho~2, Tycho~1
and Hipparcos catalogues. For each star, a scattered-light radius ($\rscat$)
was estimated based on the circular region over which the star flux per pixel
is greater than 5 times the sky noise level.  For each potential target, a
mask parameter was defined as follows
\begin{equation}
\begin{array}{l}
\mbox{{\small MASK\_IC\_12}} = 1  \\
\mbox{{\small MASK\_IC\_12}} = \rscat/d  \\
\mbox{{\small MASK\_IC\_12}} = 0
\end{array}
\mbox{~for~} 
\begin{array}{l}
d \le \rscat  \\
\rscat < d  \le 5 \rscat \\
d > 5 \rscat
\end{array}
\end{equation}
where $d$ is the distance to a $V<12$ star with radius $\rscat$.  In other
words, the {\small MASK\_IC\_12} value decreases from unity when $d \le \rscat$
to 0.2 when $d = 5 \rscat$.  A similar mask parameter {\small MASK\_IC\_10}
was defined using only $V<10$ stars. In addition, objects within an
SDSS-database mask for holes, satellite trails and bleeding pixels had these
mask values set to unity.  After testing, we chose to select only objects with
{\small MASK\_IC\_10} $< 0.5$ and {\small MASK\_IC\_12} $< 0.8$.

The largest masked areas are shown in Fig.~\ref{fig:ukidss-coverage}.  These
are between 0.01 and $0.07\,\sqdeg$ each.  Most of the separate masked areas
are significantly smaller ($<0.001\,\sqdeg$ or $<1'$ in radius).  Overall, the
total masked area is about $1.0\,\sqdeg$ and the unmasked area of the survey
is estimated to be $143.0\,\sqdeg$. 

The mask was insufficient to remove all or nearly all objects with bad
photometry.  Therefore, as per SDSS selection, objects were selected to be
\textsc{not satur} from the \textsc{flags} column in the \textsc{PhotoObj}
table. This basically excludes deblends of bright stars but will also reject
galaxies that are blended with saturated stars.  These however are likely to
have bad photometry and falsely bright magnitudes.  The stars causing this
saturation, not accounted for by the Tycho mask, are probably around $V\sim13$.

The saturated-flag masking is not ideal.  This is particularly the case for
large nearby galaxies for which the angular size of the galaxy is a
significant factor in determining the excluded sky area.  In other words, the
probability of a large galaxy having \textsc{satur} set depends primarily on
its size rather than the area of the diffracted and scattered light around
stars.  To increase the completeness of the input catalogue for large
galaxies, exceptions for the mask and not-saturated criteria were made for
galaxies from the Uppsala General Catalog (UGC; \citealt{cotton99}) and
Updated Zwicky Catalog (UZC; \citealt{falco99}).  In addition, exceptions to
the not-saturated criteria were made for a selection of visually inspected
galaxies that have \textsc{not satur\_center}. There are only 86 objects with
an exception flag set (selected as part of the visual classification process
described in \S~\ref{sec:vis-class}).

In summary, the criteria for including objects is given by:
\begin{equation}
\begin{array}{l}
  (\mbox{\textsc{mask\_ic\_10}} < 0.5 \mbox{~~\textsc{and}~~} 
   \mbox{\textsc{mask\_ic\_12}} < 0.8 \mbox{~~\textsc{and}}  \\
   \mbox{\textsc{not satur}}  )  \mbox{~~\textsc{or}~~}  
   \mbox{the exception flag is set.}
\end{array}
\label{eqn:masking}
\end{equation}

\subsection{Surface brightness limits}
\label{sec:sb-limits}

In addition to the implicit surface brightness (SB) limits from
star-galaxy separation and detection, an explicit SB limit was applied
given by
\begin{equation}
15.0 < \effsb < 26.0 
\label{eqn:sb-limits}
\end{equation}
where $\effsb$ is the effective SB in ${\rm mag\,arcsec}^{-2}$ within the 50\%
light radius in the $r$-band (eq.~5 of \citealt{strauss02}). Anything of lower
SB is very likely to be an artifact, and anything of higher SB is a star.

\begin{figure*}
\includegraphics[width=\middlecolsize\textwidth]{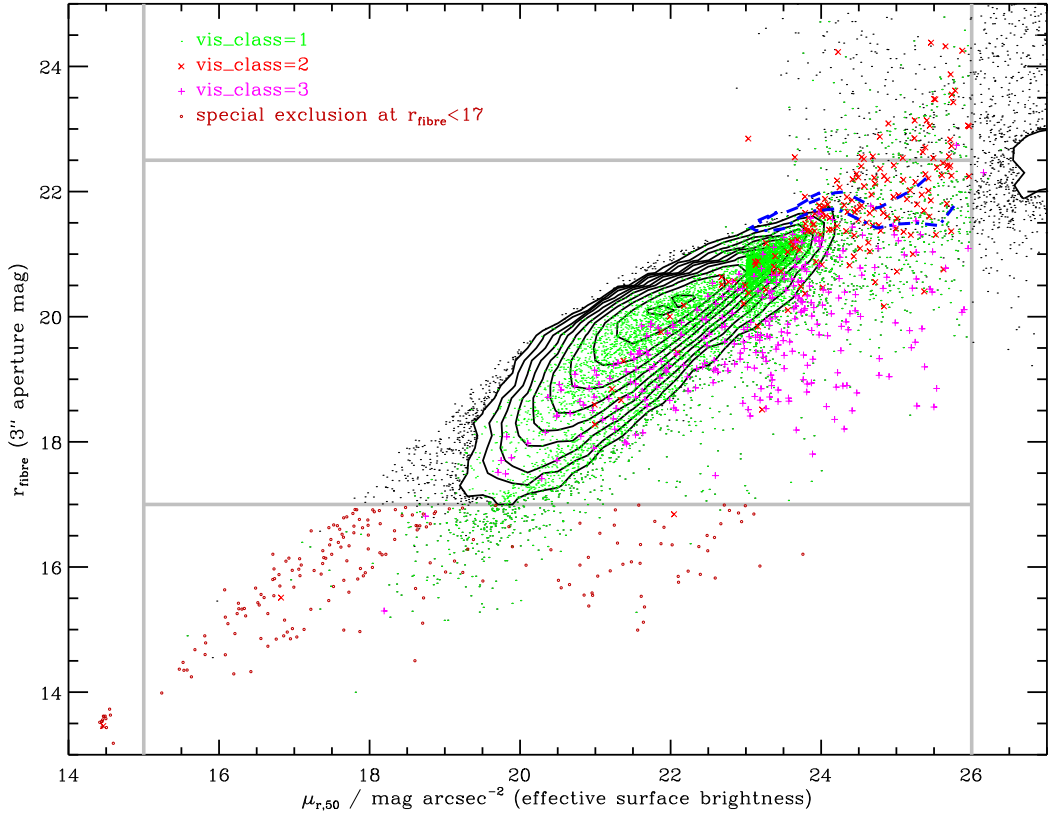}
\caption{Bivariate distribution of $r_{\rm fibre}$ versus $\effsb$.  The black
  contours and points represent objects that are not masked and pass
  star-galaxy separation.  The grey lines outline the selection limits:
  $15.0<\effsb<26.0$ is the restriction for the science catalogue; while
  objects with fibre magnitudes fainter than 22.5 or brighter than 17.0 are
  not included in the AAOmega observation schedule.  The green dots represent
  objects with \visclass=1; the red crosses \visclass=2, and the pink crosses
  \visclass=3. The small red circles at $r_{\rm fibre}<17$ are probably stars
  based on a stricter star-galaxy separation criteria
  (\S~\ref{sec:bright-galaxies}). The blue dash-dotted (dashed) line
  corresponds to 50\% (30\%) redshift success rate for objects on or near the
  line.}
\label{fig:sb-sep}
\end{figure*}

Figure~\ref{fig:sb-sep} shows the distribution of objects in $r_{\rm fibre}$
versus $\effsb$ for GAMA main-survey targets. The lower limit of 15.0 does
remove some objects, probably stars, not rejected by the masking or
star-galaxy separation criteria (Eq.~\ref{eqn:star-gal-summary}).  The limit
for $\effsb$ of $26.0$ is 1.5~magnitudes deeper than the SDSS MGS cut,
and is the point at which most of the objects are clearly artifacts.  Note
that the SDSS photometric pipeline is not complete for $\effsb > 23$
(figs.~2--3 of \citealt{blanton05}).  Additional low-SB candidates could be
recovered by searching coadded $g$, $r$ and $i$ images \citep{kniazev04}.
Nevertheless without deeper imaging, the data will remain incomplete at low SB
well before our explicit limit.

In addition to the explicit SB limits given in
Eq.~\ref{eqn:sb-limits}, which we use to reject objects from our
science catalogue, we include a restriction on the fibre magnitudes:
\begin{equation}
17.0 < r_{\rm fibre} < 22.5 
\label{eqn:fibre-limits}
\end{equation}
for targets allocated to the AAOmega observation schedule.  This is a
practical restriction, with a bright limit to avoid significant crosstalk in
the spectrograph and a faint limit because the redshift success is very
low. Selected fibre bright targets without a known redshift will be observed
with a 2m-class telescope, and, in principle, selected fibre faint targets
will be observed with an 8m-class telescope. At the bright end, a more
restrictive cut on star-galaxy separation is also justified (see later in
\S~\ref{sec:bright-galaxies}).

\subsection{Visual classification}
\label{sec:vis-class}

Sources with, for example, $\effsb > 23$ have a high probability of being
artifacts, deblends of stars, or the outer parts of galaxies.  One of us
(J.\ Liske) has written code to facilitate the visual classification of such
sources. A \visclass\ variable, initially with zero value, could be changed to
the following for each source on inspection:
\begin{itemize}
\item [1] possibly a target, 
\item [2] not a target (no evidence of galaxy light), 
\item [3] not a target (not the main part of a galaxy).
\end{itemize}

First, sources with the following flags all equal to zero, \textsc{edge,
  blended, child, maybe\_cr, maybe\_eghost}, were assumed to be good,
essentially isolated, and not included in any testing (\visclass\ set to 255).
About 50\% of targets satisfy these criteria.  From the remaining objects,
sources were selected for visual classification if any of the following
conditions applied: $\effsb > 23$, $r_{\rm fibre} > 21$, $r_{\rm fibre} < 17$,
\textsc{mask\_ic\_12} $> 0.2$, $r_{\rm model} < 15.5$, $r_{\rm petro} < 15.5$,
$r_{\rm fibre} < r_{\rm model}$, $r_{\rm fibre} < r_{\rm petro}$, near UGC
galaxy, within $3''$ of another target, Petrosian radius $>10''$.  These
indicate that the object could be the result of deblending of a large galaxy,
artifact or bright star, e.g., diffraction spikes.  In addition to the above
criteria, other objects were included in the above process.  Objects with the
same \textsc{parentid} as an already classified \visclass\ = 3 object were
selected. (The above selection was not developed in one go and there have been
several iterations.) Finally objects, with the same \textsc{parentid}, that are
the brightest and nearest to any object to be tested were included. Objects
that could be part of the same galaxy were viewed together where possible. One
had to be certain to classify objects as 3 only if the main part was
identified as a target.

The above selection produced a sample of about 12\,500 objects for visual
classification, by six observers. Every selected object was classified by
three different observers.  Of the selected potential main-survey targets
(Fig.~\ref{fig:sb-sep}), \visclass=1 was set in 92\% of cases, \visclass=2 in
5\% of cases, and \visclass=3 in 3\% of cases, based on agreement between two
or all three classifiers, 9\% and 90\% of cases, respectively.  Some of the
ambiguous cases were double checked, and a single-observer classification was
selected in 1\% of cases.  Objects with values of 2 or 3 were removed from the
schedule of AAT observations, i.e., targets must satisfy
\begin{equation}
\mbox{\visclass} \neq 2 \mbox{~~\textsc{and}~~} \mbox{\visclass} \neq 3 \: .
\label{eqn:vis-class}
\end{equation}
In addition, the $\visclass=3$ objects can be used to improve the photometry
of some large galaxies by coadding in the flux of the galaxy parts (or the
`parent' photometry can be used).

\subsection{Number of targets}
\label{sec:number-targs}

The total number of objects that are within the GAMA regions
(\S~\ref{sec:sdss}), main-survey magnitude limits (Eq.~\ref{eqn:mag-limits})
and $\dsg > 0.05$, is 143\,728. Applying the stricter star-galaxy separation
(Eq.~\ref{eqn:star-gal-summary}) reduces the sample to 132\,073.  Removing
objects by masking (Eq.~\ref{eqn:masking}), the SB limits
(Eq.~\ref{eqn:sb-limits}) and visual checking (Eq.~\ref{eqn:vis-class}),
reduces the sample to 120\,038. Of these, 825 were not included in the AAOmega
observation schedule because they do not satisfy the fibre magnitude limits
(Eq.~\ref{eqn:fibre-limits}).  A more restictive star-galaxy separation can be
applied for brighter targets (discussed later and given in
Eq.~\ref{eqn:star-gal-special}) that reduces the sample to {\bf 119\,852}.
This is considered to be the main-survey sample.  Note these numbers apply to
AAT observations in 2009, the numbers may change slightly with addition of
complete $J$-$K$ UKIDSS.

Separating the main survey into $r$, $z$ and $K$ limited samples, the
numbers are 114\,520, 61\,418 and 57\,657, respectively.  \newtext{For
  $r_{\rm petro}$ selected samples to 19.0, 19.4 and 19.8, the sample
  sizes are 60\,407, 96\,386 and 150\,810, respectively.  The latter
  is the $r$-selected main survey plus F2 additional targets, which
  are described in the following section.}

\subsection{Additional targets}
\label{sec:filler-targets}

In order to assess the spectro-photometry of the AAOmega spectra, three or
four stars, classified as \textsc{redden\_std} or \textsc{spectrophoto\_std}
by SDSS, were observed in each configuration. These also had a bright fibre
magnitude limit of 17 as per the main-survey targets.

The aim is to obtain high completeness (99\%), at least in terms of spectra
obtained and ideally in terms of confirmed redshifts, for the main survey.
This is set to reduce systematic uncertainties in GAMA's position dependent
science cases, and is possible because a given patch of sky is potentially
observed by $\sim5$--10 2dF tiles depending on the local density of targets
(see \citealt{robotham09} for a description of the tiling strategy).  Given
this requirement, targeting becomes increasingly inefficient as the survey
progresses (fewer targets without a redshift per tile). Filler targets were
introduced to provide useful redshifts outside the main survey, and thus,
maximise fibre usage.  These have no high-level requirement on
completeness. The filler selections are given by: (F1) objects with detection
in the Faint Images of the Radio Sky at Twenty-cm (FIRST) survey and matched
to SDSS with $i_{\rm model} < 20.5$ including unresolved sources; (F2) $19.4 <
r_{\rm petro} < 19.8$ galaxy targets in G09 and G15, aiming for equal depth
with G12; and (F3) $g_{\rm model}<20.6$ or $r_{\rm model}<19.8$ or $i_{\rm
  model}<19.4$ in G12, investigating variation in magnitude-type and
wavelength on selection. In total, there are about 50\,000 filler targets.

\section{Spectroscopy}
\label{sec:spec}

\subsection{Existing data sets}
\label{sec:exist-spec}

While the GAMA target density is significantly higher than SDSS or 2dFGRS,
the redshifts obtained by these and other surveys provide a non-negligible
starting baseline. We incorporate a number of different surveys into our
catalogue, defining a redshift quality $Q$, where necessary, as per the
\citet{colless01} scheme such that $Q=1$ means very poor or no
redshift, $Q=2$ means a possible but doubtful redshift, $Q=3$ means a probable
redshift, and $Q=4$ or $Q=5$ means a reliable redshift. The surveys included
are given in Table~\ref{tab:z-surveys-details}.

\begin{table*}
\caption{Other spectroscopic data in the GAMA regions. Tables were obtained
  from the survey websites or the VizieR service.}
\label{tab:z-surveys-details}
\begin{tabular}{lllrrr} \hline
survey  & file/table              & reference       & 
  no.\ of redshifts\rlap{$^a$} & no.\ $Q\ge3$  & no.\ main survey unique\rlap{$^b$} \\ \hline
SDSS     & DR7 \textsc{specobjall} &\citet{sdssDR7}   & 27514 &  26687\rlap{$^c$} & 13170 \\
2dFGRS   & VII/250/2dfgrs          &\citet{colless03} & 11490 &  11180            &  3840 \\
MGCz     & VII/240/mgczcat         &\citet{driver05}  &  4008 &   3835            &  1883 \\
2SLAQ-LRG& J/MNRAS/372/425/catalog &\citet{cannon06}  &  2256 &   2109            &   227 \\
6dFGS    & DR3 \textsc{spectra}    &\citet{jones09}   &   299 &    270            &    55 \\
UZC      & J/PASP/111/438/catalog  &\citet{falco99}   &   255 &    209\rlap{$^d$} &    13 \\
2QZ      & VII/241/2qz             &\citet{croom04}   &  5359 &   4317\rlap{$^e$} &   224 \\
2SLAQ-QSO& 2slaq\_qso\_public.cat  &\citet{croom09}   &  2414 &   2098\rlap{$^e$} &    34 \\ \hline
\end{tabular}
\begin{flushleft}
$^a$The number of redshifts quoted are all those in the GAMA
regions including duplicates and non-GAMA targets.  \\
$^b$The number corresponds to unique main survey targets with
  a $Q\ge3$ redshift from the survey (prior to GAMA). 
  In the case of multiple matches within $1''$, the highest $Q$ value
  match is used (nearest in case of equal $Q$):
  $Q$ is limited to $\le4$ for all surveys except SDSS. \\
$^c$SDSS quality is given by $Q = 1 + (\textsc{zconf} > 0.2) +
(\mbox{zwarning\_okay} \textsc{~and~} \textsc{zconf} > 0.7) + (\textsc{zconf}
> 0.9) + (\textsc{zconf} > 0.99)$ where each term in brackets takes the value 
of unity if the condition is true and zero otherwise, and zwarning\_okay takes 
the value unity if the following warning flags 
\textsc{emab\_inc, ab\_inc, 4000break} are all zero. \\ 
$^d$UZC quality is given by: $Q=3$ if UZC class is 0 or 1 (secure
identification), and $Q=2$ if UZC class is 2, 3 or 4 (some confusion regarding
identification).  \\ 
$^e$2QZ and 2SLAQ-QSO quality is given by: $Q=3$ if original quality code was 11
(good identification and redshift); $Q=2$ if 22, 12 or 21; and $Q=1$ if 33, 23
or 32. 
\end{flushleft}
\end{table*}

From the $Q\ge3$ non-GAMA redshifts in the GAMA regions as outlined in the
table, about 40\,000 are unique (considering matches within $1''$ to be the
same object). The number of main survey targets with one of these redshifts is
19\,446, matching within $1''$ except for some large galaxies within $3''$ of
a 6dFGS or UZC redshift. \newtext{The non-GAMA redshifts without a match to a
  GAMA target are primarily of stars and quasars.}  Objects with $Q\ge3$
redshifts are given a lower priority in the AAOmega observation schedule.

\subsection{AAOmega observations in 2008 and 2009}
\label{sec:aaomega}

GAMA observations with the multi-object spectrograph AAOmega on the AAT took
place in 2008 (Jan\,12, Feb\,29 to Mar\,15, Mar\,30 to Apr\,05) and 2009
(Feb\,27 to Mar\,05, Mar\,27 to Apr\,02, Apr\,17 to Apr\,23).  The 2dF robotic
fibre positioner \citep{lewis02df} feeds a bench-mounted dual-beam
spectrograph \citep{sharp06}.  Two plates are used: while one is being
configured (fibres placed), the other plate is in the focal plane feeding
light to the spectrograph.  There are up to 392 science fibres available in a
single configuration.  Excluding broken fibres, 20--25 fibres used for sky
subtraction and 3 or 4 spectroscopic standards (\S~\ref{sec:filler-targets}),
we targeted between 320 and 350 GAMA targets per configuration. Total exposure
times used were typically 1~hour ($3\times20{\rm\,min}$). We observed up to 8
configurations in a single night for a total of 267 observations over the two
years (91\,015 spectra). The spectral coverage was from 370 to 880\,nm.

The priorities assigned to targets were different between the two years.  
The tiling scheme is described in detail by \citet{robotham09}.  Here, we
summarize the priorities.  In 2008 the targets consisted only of the $r$-band
selection with $\dsg>0.25$ (there was insufficient UKIDSS coverage at the
time), without an already known redshift (\S~\ref{sec:exist-spec}) except for
some cross-check data.  The priorities were from high-to-low: (i) $r<19.0$;
(ii) $19.0<r<19.8$ in G12 within $\pm0.5\degr$ of the celestial equator; (iii)
$19.0<r<19.4$ in G09 and G15, and remaining $19.0<r<19.8$ in G12.  In
addition, clustered targets in any of these categories were given a higher
priority. A clustered target was defined as one within $40''$ of another
target, where $40''$ is approximately the closest two fibres can be
placed. This was to maximise the chances of observing as many close pairs as
possible over three years of observations.  In 2009, now including UKIDSS
selection for the star-galaxy separation and magnitude limits, the priorities
were: (i) clustered unobserved main-survey targets; (ii) unobserved main
survey or clustered failed main survey, where failed means that a GAMA
spectrum has been obtained with $Q\le2$; (iii) failed main survey; (iv) from
F1, F2, F3 filler targets, and $Q=3$ spectra taken with the old 2dF
spectrographs (e.g., 2dFGRS).

From the first two years of observing, first-pass reductions with
\textsc{2dfdr} \citep{CSH04} and \textsc{runz} \citep{SCS04} have resulted in
a 94 per cent redshift success rate ($Q\ge3$) for 82\,696 unique redshifts,
80\,944 for the main survey (79\,599 with
$z>0.002$). Table~\ref{tab:gama-spec} gives a breakdown of the spectra
obtained.  Including spectra from other surveys, results in 100\,012 $Q\ge3$
redshifts for the main survey (98\,497 with $z>0.002$).
Table~\ref{tab:gama-main-spec} gives the target numbers and redshift
completeness for various main survey selections.  \newtext{Note particularly
  the drop in completeness between $r_{\rm petro}<19.0$ (96\% average
  completeness) and the fainter $r$-band selection (74\%), and a further drop
  to the $z$- and $K$-band extra selection (39\%).  The $r$-limit only
  selection and the prioritisation in the first year is the main cause of
  differing $Q\ge3$ completeness factors between each sub-sample, i.e., it is
  primarily a variation in targeting completeness though redshift success rate
  is also lower for the fainter samples. No observations based on $J$ and/or
  $K$-band photometry were started in the first year so the marginally
  resolved sample within each magnitude range is also of lower completeness.}

\begin{table}
\caption{GAMA spectra from AAT observations in 2008 and 2009}
\label{tab:gama-spec}
\begin{center}
\begin{tabular}{lr} \hline
description                     &  number \\ \hline
total spectra obtained          &  91015  \\
~~~spectroscopic standards      &   1059  \\
~~~unique targets               &  87753  \\
~~~repeated targets             &   2203  \\ \hline
$Q\ge3$ unique targets$^a$      &  82696  \\
~~~$r<19.0$ \& $\dsg>0.25$         &  40103  \\
~~~main survey $r$-selected        &  38994  \\
~~~main survey $z,K$-selected      &   1847  \\
~~~F1: radio selected              &    105  \\
~~~F2: $19.4<r<19.8$ in G09 \& G15 &   1029  \\
~~~F3: filler selection in G12     &     68  \\
~~~other$^b$                       &    550  \\ \hline
\end{tabular}
\end{center}
$^a$The unique targets with redshifts are identified in the rows below.  The
$r<19.0$ selection corresponds to the higher priority targets in the first
year of AAT observations. The numbers shown in each row below this row do not
include contributions already accounted for. Below the main survey are the
F1--F3 filler targets (\S~\ref{sec:filler-targets}).  \\ 
$^b$The `other' objects are mostly objects whose UKIDSS photometry has
undergone revision since the second year of AAT observations, and \visclass=3
objects that were observed prior to implementation of the visual
classification.
\end{table}

\begin{table*}
\caption{Main survey target numbers and redshift completeness for the three
  separate GAMA regions, galaxy fractions (from $Q\ge3$ redshifts), and median
  galaxy redshifts.  The redshift completeness is defined as the number of
  objects with a $Q\ge3$ redshift divided by the number of targets (regardless
  of whether they have been observed spectroscopically).}
\label{tab:gama-main-spec}
\begin{center}
\begin{tabular}{lrrrrrrrr} 
\hline
selection & \multicolumn{2}{c}{Region G09} & \multicolumn{2}{c}{Region G12} & \multicolumn{2}{c}{Region G15}
& fraction & median \\
          & no. targets & $Q\ge3$   & no. targets & $Q\ge3$   & no. targets & $Q\ge3$   & $z>0.002$&redshift\\
\hline
$r_{\rm petro}<16.0$                       &    363&  95.0\% &     397&  97.5\% &     481&  96.9\% &  99.1\% &   0.052\\
$16.0<r_{\rm petro}<17.8$                  &   3335&  99.2\% &    4644&  99.3\% &    4666&  99.0\% &  98.5\% &   0.116\\
$17.8<r_{\rm petro}<19.0$ \& $\dsg>0.25$   &  14387&  98.2\% &   15599&  96.1\% &   16016&  93.1\% &  99.0\% &   0.185\\
$17.8<r_{\rm petro}<19.0$ \& $\dsg<0.25$   &    160&  83.8\% &     206&  60.7\% &     153&  55.6\% &  66.3\% &   0.260\\
$19.0<r_{\rm petro}<19.4$ \& $\dsg>0.25$   &  11886&  90.6\% &   11600&  76.4\% &   11724&  62.6\% &  98.9\% &   0.243\\
$19.0<r_{\rm petro}<19.4$ \& $\dsg<0.25$   &    201&  77.1\% &     345&  46.1\% &     223&  33.6\% &  80.7\% &   0.228\\
$19.4<r_{\rm petro}<19.8$ \& $\dsg>0.25$   &    ---&   ---   &   17281&  70.1\% &     ---&   ---   &  99.5\% &   0.263\\
$19.4<r_{\rm petro}<19.8$ \& $\dsg<0.25$   &    ---&   ---   &     853&  38.9\% &     ---&   ---   &  93.7\% &   0.254\\
$z_{\rm model}<18.2$ and not $r$-selected  &    604&  63.1\% &     270&  25.2\% &     510&  38.2\% &  58.4\% &   0.470\\
$K_{\rm AB,auto}<17.6$ and not $rz$-selected&  1931&  46.5\% &     348&  19.0\% &    1669&  29.5\% &  97.3\% &   0.368\\
\hline
all main survey                         &   32867&  91.6\% &   51543&  80.9\% &   35442&  79.5\% &  98.5\% &   0.196\\ 
\hline
\end{tabular}
\end{center}
\end{table*}

The details of spectroscopic data reduction, including new defringing and
sky-subtraction techniques, redshifting, comparison with other spectra,
spatial and magnitude completeness will be described in future GAMA papers.
In the next section, we use the first-pass redshifts to illustrate some issues
related to the target selection.

\section{Results}
\label{sec:results}

\subsection{Star-galaxy separation}
\label{sec:results-sg-sep}

\begin{figure*}
\includegraphics[width=\middlecolsize\textwidth]{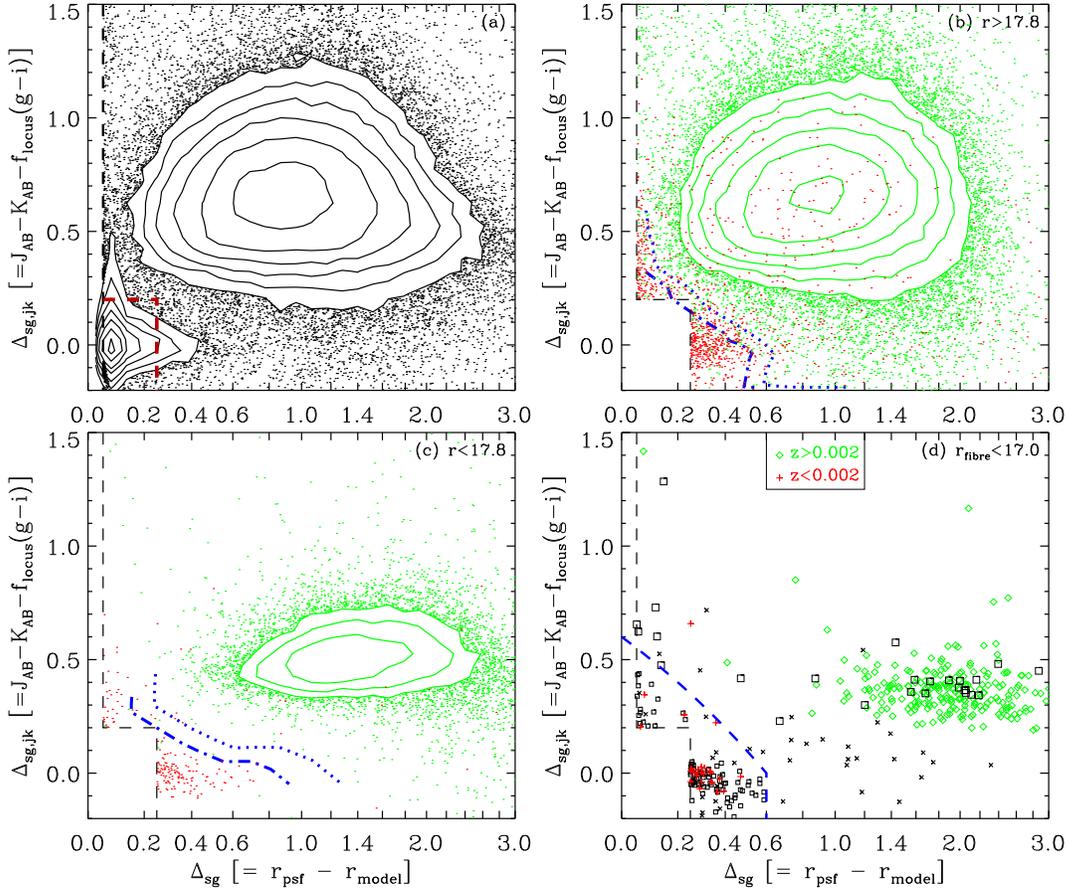}
\caption{Results of star-galaxy separation. {\bf (a):} The distribution of
  main survey targets with a $J-K$ measurement, extended to all objects with
  $\dsg>0.05$, are shown with black contours and points.  The red dashed line
  shows the cut used for target selection. {\bf (b,c):} The distribution of
  objects ($r_{\rm petro}>17.8,<17.8$) with confirmed galaxy redshifts are
  shown with green contours and points, while objects with stellar redshifts
  are shown with red points.  The blue dotted (dash-dotted) line corresponds
  to 50\% (70\%) stellar contamination for objects on or near the line
  (determined by interpolation in the $r<17.8$ sample). {\bf (d):} Objects
  with $r_{\rm fibre}<17.0$, not included in the AAOmega observation schedule,
  are shown. Red crosses and green diamonds represent objects with confirmed
  stellar and galaxy redshifts.  Black crosses (squares) represent objects
  where the fibre magnitude is brighter (fainter) than the Petrosian
  magnitude. The smaller squares and crosses are the potential targets
  excluded by the criteria of Eq.~\ref{eqn:star-gal-special}: these are also
  shown as small red circles in Fig.~\ref{fig:sb-sep}.  The blue dashed
  line divides the small and large squares.}
\label{fig:results-sg-sep}
\end{figure*}

There are two star-galaxy separation parameters used in the GAMA
selection. Figure~\ref{fig:results-sg-sep}(a) shows the observed bivariate
distribution of main survey targets in these parameters.  The red line shows
the cut used for our target selection.  This removes nearly 9\,000 sources or
about 7\% of potential targets to $\dsg>0.05$.
Figure~\ref{fig:results-sg-sep}(b,c) show the distributions of galaxies
($z>0.002$) and stars that have confirmed redshifts, 1.5\% are stellar, using
all available spectroscopic data.  The additional $J-K$ selection was
necessary for sources with $r>17.8$ in order to be complete for compact
galaxies.  This is seen by the confirmed galaxy contours extending to the
left of $\dsg=0.25$ in Fig.~\ref{fig:results-sg-sep}(b), which would
otherwise have been missed by using only a $\dsg>0.25$ cut.  Note that the
targeting completeness is lower at $\dsg<0.25$, 60\% compared to nearly 90\%
overall, because this UKIDSS-SDSS selection was not available for AAT
observations in 2008.

Figure~\ref{fig:results-sg-sep} also shows that the regions of high stellar
contamination are, not surprisingly, at low $\dsg$ or $\dsgjk$. Thus a lower
contamination could be obtained by using a cut $\dsg+\dsgjk > 0.4$, for
example, with minimal rejection of genuine galaxies.  This would work well
because there is no strong correlation between the two parameters.

An estimate of the completeness of the current selection in terms of selecting
galaxies can be obtained by assuming that there is no significant correlation
between $\dsg$ and $\dsgjk$.  Consider the galaxy distribution in
Fig.~\ref{fig:results-sg-sep}(b).  The fraction of galaxies at $\dsgjk<0.2$ is
2.3\% (not including galaxies with no $J-K$ measurement) and the fraction at
$\dsg<0.25$ is 1.7\% after adjusting the latter for the lower targeting
completeness. Thus the predicted fraction of galaxies at $\dsgjk<0.2$ and
$\dsg<0.25$ (in the lower-left hand corner of the plot) is only 0.04\%.  Thus,
the galaxy selection from the star-galaxy separation is plausibly $\ga99.9$\%
complete when there are $J$ and $K$ measurements. This assumes there is no
significant population of galaxies with $\dsg<0.05$ within our magnitude
limits.

\begin{figure}
\centerline{
\includegraphics[width=\singlecolsize\textwidth]{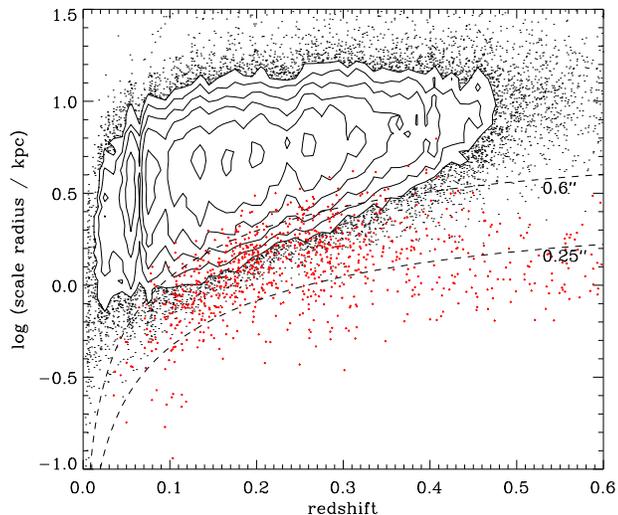}
}
\caption{Scale radius, de Vaucouleurs or exponential profile from
  \textsc{photo}, versus redshift. The black contours and points
  represent galaxies selected using $\dsg>0.25$, while red crosses
  represent the $J-K$ selected sample with $0.05<\dsg<0.25$. The
  dashed lines correspond to constant observed angular size.}
\label{fig:size-z}
\end{figure}

The SDSS pipeline \textsc{photo} also determines the scale radii of the de
Vaucouleurs and exponential profile fits (eq.~9 \&~10 of
\citealt{stoughton02}).  Taking the best fit and averaging the scale radii in
the $r$- and $i$-bands for each galaxy, we determined the completeness in this
measure of size.  The cut $\dsg>0.25$ is complete down to a scale radius
$\sim0.6''$, while our star-galaxy separation (Eq.~\ref{eqn:star-gal-summary})
is plausibly complete down to a scale radius $\sim0.25''$.\footnote{We note
  that the \textsc{photo} scale radius values should be interpreted with some
  caution at small sizes, less than half the typical PSF
  width. \citet{taylor09} advocate treating objects with scale radii $<0.75''$
  as having an upper limit of $0.75''$, i.e., the true value is poorly
  determined even though \textsc{photo} has determined that the object is
  likely to be resolved.}  Figure~\ref{fig:size-z} shows the scale radius in kpc
versus redshift for confirmed galaxies in the main survey. Without the
additional selection, the target selection would be significantly incomplete,
$\sim20$\% missed, for galaxies with observed radii between $0.25''$ and
$0.6''$. Of course, one could have used this scale radius directly as a
star-galaxy separation parameter but, without higher resolution imaging, the
systematic errors are presumably larger in this than $\dsg$.

This compact galaxy selection is critical for studies with a direct interest
in the size evolution of galaxies (e.g.,
\citealt{trujillo06,CD07,taylor09}). Targeting all objects with $\dsg>0.05$
would have resulted in $\sim9\,000$ extra objects, which would have been a
very inefficient way to target compact galaxies.  Future higher S/N and higher
resolution imaging (optical and near-IR) will improve the efficiency of this
type of selection, providing a test of whether GAMA target selection has
missed significant numbers of compact galaxies.

\subsection{Bright galaxies}
\label{sec:bright-galaxies}

Objects with $r_{\rm fibre}<17.0$ are not allocated to the AAOmega schedule to
avoid crosstalk between spectra tramlines, and we do not need to consider
these objects for this target selection. However, it is necessary for analyses
at low redshift, e.g.\ measuring luminosity functions, to determine a
realistic completeness of galaxy selection at bright magnitudes.
Figure~\ref{fig:results-sg-sep}(d) shows the distribution in the star-galaxy
separation parameters for these potential targets.  There are 485 using our
normal selection criteria, of which, 296 have redshifts from SDSS and other
surveys ($Q\ge3$; Table~\ref{tab:z-surveys-details}). One possibility would be
to observe all remaining 189 targets with a 2m-class telescope. However, most
of these are probably stars and a more restrictive criterion could be used.
This is given by
\begin{equation}
\begin{array}{ll} 
  r_{\rm fibre} > 17.0 & \mbox{\textsc{or}} \\
  (\dsg+\dsgjk > 0.6 \mbox{~~\textsc{or}~~} \dsg > 0.6 & \mbox{\textsc{and}} \\
  r_{\rm fibre} > r_{\rm petro}) \: .
\end{array}
\label{eqn:star-gal-special}
\end{equation}
The $\dsg$-$\dsgjk$ cut is shown by the blue dashed line in
Fig.~\ref{fig:results-sg-sep}(d), while targets that satisfy the last criteria
are shown as squares as opposed to crosses. Sources with fibre magnitude
brighter than Petrosian are indicative of a `possible' galaxy blended with a
star, however, the star light dominates the fibre magnitude, which is not
deblended. Using the above cut results in 299 sources with 266 redshifts (89\%
complete).  This cut should be used when assessing completeness at the bright
end of GAMA targets.  This was applied before computing the $r_{\rm petro}<16$
target numbers and completeness given in Table~\ref{tab:gama-main-spec}.

\subsection{Low surface brightness galaxies}
\label{sec:low-sb-galaxies}

The completeness in the low SB regime depends on redshift success and source
detection \citep{DP83,blanton05}, and there is the additional issue of the
accuracy of the flux measurements \citep{CD07}. These will be described in
detail in a future paper on luminosity functions (Loveday et al.\ in
preparation).  Here we note only that the redshift success rate is primarily a
function of $r_{\rm fibre}$ as shown in Fig.~\ref{fig:sb-sep}. The success
rate is 50\% at $r_{\rm fibre} \sim 21.5$. This does not include any coadding
of GAMA spectra over two or more observations, and there may be improvement
after re-reduction.

\subsection{Redshift distributions and near-IR selections}
\label{sec:z-distribution}

\begin{figure}
\centerline{
\includegraphics[width=\singlecolsize\textwidth]{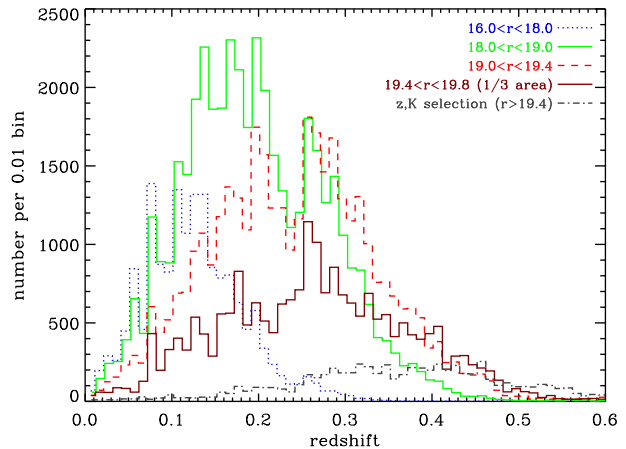}
}
\caption{Redshift distributions for selected main-survey samples.
  The numbers have been projected to completion of the main survey 
  using empirical completeness determined in $g-i$ bins.}
\label{fig:z-histos}
\end{figure}

Not accounting for incompleteness, 50\% of the galaxy redshifts are in the
range 0.13--0.27, 90\% are in the range 0.06--0.39 and 99\% are in the range
0.02--0.53. Figure~\ref{fig:z-histos} shows the redshift histograms for
various galaxy samples ($z>0.002$) within the main survey, and median
redshifts are given in Table~\ref{tab:gama-main-spec}.  The near-IR selections
have a higher average redshift. Note that the redshift distribution within
each sub-sample may be biased by non-GAMA redshifts and the dependence of
redshift success rate on magnitude, for example. These are corrected for in
Fig.~\ref{fig:z-histos} by binning in $g-i$ to determine completeness
factors. The histogram for each sub-sample is determined by weighting each
object with a redshift by $1/c$ where $c$ is the redshift completeness in each
bin (with bin size of 0.2 for $0<g-i<3$). This colour is used because of its
correlation with redshift [Fig.~\ref{fig:ukidss-sg-sep}(b)].
 
\begin{figure}
\centerline{
\includegraphics[width=\singlecolsize\textwidth]{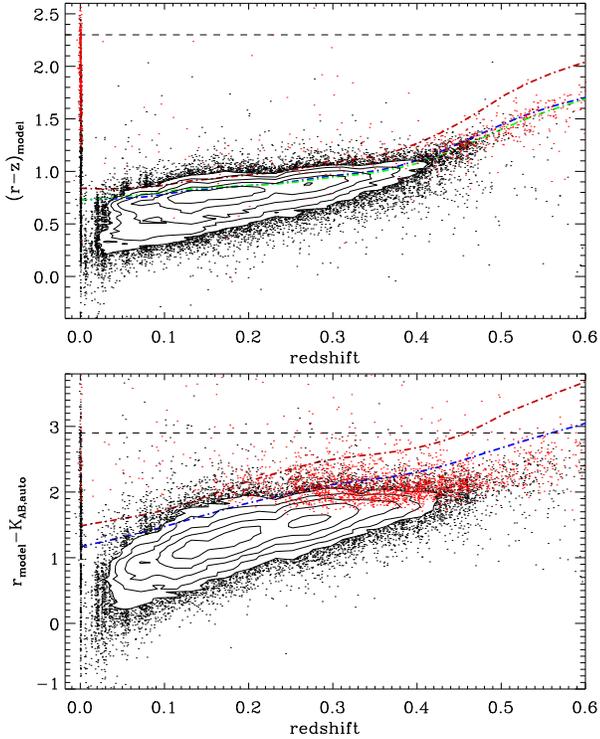}
}
\caption{Observed colour versus redshift for (a) the $z_{\rm model}<18.2$
  sample and (b) the $K_{\rm AB,auto}<17.6$ sample. The black contours and points
  represent the data within $r_{\rm petro}<19.4$ (or $z_{\rm model}<18.2$
  for the $K$-selected sample), while the red points represent the remaining
  fainter selection. The dash-dotted lines represent the observed colours of
  SSP models with $z_{\rm form}=6$ (12.5\,Gyr old at $z=0$, 7.5\,Gyr at
  $z=0.5$): red ($Z=0.05$, \citealt{bc03}), blue ($Z=0.04$, alpha-enhanced
  abundances, \citealt{percival09}), and green ($Z=0.02$, empirical stellar
  spectra, \citealt{maraston09}).  The horizontal dashed line represents the
  completeness limit at the faint end of the samples given the $r_{\rm
    model}<20.5$ limit (Fig.~\ref{fig:color-bias}).}
\label{fig:obs-color-z}
\end{figure}

Figure~\ref{fig:obs-color-z}(a) shows observed $r-z$ versus redshift for the
$z$-selected sample, with the targets fainter than 19.4 in $r$ shown by red
points. The extra $z$-band selection is mostly picking up luminous galaxies in
the redshift range 0.4--0.6 (recalling that this is of lower completeness than
the $r<19.4$ selection). The number density of targets drops off well before
the colour bias limit.  Simple stellar population (SSP) tracks are shown with
a formation redshift of six (see caption for references). Some objects are
apparently redder than the old SSP tracks. This is presumably mostly because
of photometric errors, however, certain dust geometries can in principle
redden galaxies beyond the colour of old stellar populations.  Dusty galaxies
can lie on, and slightly redder than, the red sequence \citep{WGM05}.  Note
that most of the targets in the range $1.2 < r-z < 2.5$ are a stellar
contamination \newtext{(or 41.6\% of the $z$-band extra selection,
  see Table~\ref{tab:gama-main-spec})}.

Figure~\ref{fig:obs-color-z}(b) shows observed $r_{\rm model}-K_{\rm AB,auto}$
versus redshift for the $K$-selected sample, with the targets fainter than
19.4 in $r$ and 18.2 in $z$ shown by red points. The extra selection is mostly
picking up red galaxies in the redshift range 0.2--0.5. The tracks show that
the $K$ selection is possibly incomplete for maximally old super-solar
metallicity populations at redshift $>0.45$ (from one of the models). There
are many sources significantly redder than the tracks, however, this is most
probably because of the mismatch in apertures between the surveys (model versus
\textsc{auto} mags, different deblending algorithms). For most purposes, it
would be adequate to assume the selection is $K$-band limited only.

\section{Summary}
\label{sec:summary}

The GAMA survey is designed to be a highly complete redshift survey with a
target density several times that of SDSS. The survey covers three
$48\,\sqdeg$ regions near the celestial equator centred on 9\,h, 12\,h and
14.5\,h (Fig.~\ref{fig:sdss-stripes}).  The input catalogue is drawn from the
SDSS and UKIDSS.  The main-survey limits are $r_{\rm petro}<19.4$, $z_{\rm
  model}<18.2$ and $K_{\rm AB,auto}<17.6$ ($K<15.7$) across all the regions,
and $r_{\rm petro}<19.8$ over the G12 region (Eq.~\ref{eqn:mag-limits}).  This
corresponds to a main survey of 119\,852 targets.  The near-IR selections have
a joint constraint with $r_{\rm model}<20.5$, which has minimal impact on the
use of the near-IR selections (Figs.~\ref{fig:color-bias}
\&~\ref{fig:obs-color-z}). The GAMA survey lies between that of the SDSS-MGS
$r<17.8$ and VVDS-wide $I_{\rm AB}<22.5$ magnitude-limited samples in the
depth-area plane (Fig.~\ref{fig:compare-grs}).  In terms of $K$-band selection
(Figs.~\ref{fig:ukidss-coverage} \&~\ref{fig:K-counts}), GAMA covers an area
$\sim15$ times that of the similar-depth Hawaii+AAO $K<15$ survey.

In order to be highly complete at the high-SB end of the galaxy distribution,
an intensity profile parameter (Eq.~\ref{eqn:sdss-star-gal-sep}) and a
colour-colour parameter (Eq.~\ref{eqn:ukidss-sdss-star-gal-sep}) are used
jointly for star-galaxy separation.  The $\dsgjk$ parameter makes use of $J-K$
and $g-i$ colours.  Either parameter works reasonably well in separating stars
and galaxies (Figs.~\ref{fig:sg-sep-histo}--\ref{fig:test-auto-jk}). A joint
selection (Eq.~\ref{eqn:star-gal-summary}) increases the completeness while
stellar contamination in the sample remains at less than 2\%.  Judging by the
joint distribution of confirmed galaxies in these parameters
(Fig.~\ref{fig:results-sg-sep}), the completeness is high because the
bivariate density drops significantly prior to the limit of our
selection. This is particularly important when considering the size evolution
of galaxies (Fig.~\ref{fig:size-z}). The incompleteness at the low-SB end is
significant, both in source detection and redshift success rate, which is
about 50\% at $r_{\rm fibre}=21.5$ (Fig.~\ref{fig:sb-sep}).  Some improvement
over the SDSS MGS is made by visually checking low-SB targets ($\effsb > 23$),
rather than using automatic checks, by increased redshift success rate, and by
eventually including further integrations of sources with failed redshifts.

The GAMA survey has completed two out of a three-year time allocation for
spectroscopy with AAOmega on the AAT. To date, 100\,012 redshifts have been
confirmed for the main survey, including 80\,944 from AAOmega.  Of these, 98.5
per cent are extragalactic.  \newtext{The completeness is 96\% for $r_{\rm
    petro}<19.0$, 74\% for the fainter $r$-band selection, and $\sim39$\% for
  the remaining near-IR selection (Table~\ref{tab:gama-main-spec}).  The
  completeness at $r>19$ will be significantly improved in the third year of
  spectroscopic observations.}  We expect that this galaxy redshift survey
will form a core of a fundamental database for many studies in extragalactic
astronomy.

\section*{Acknowledgements}

This part of the GAMA survey has been made possible by the efforts of staff at
the Anglo-Australian Observatory, and members of the Sloan Digital Sky Survey
and UKIRT Infrared Deep Sky Survey teams. We thank Steven Warren in particular
for efforts in making the GAMA regions a high priority in UKIDSS, the
anonymous referee for useful suggestions, and Daniel Mortlock for discussion
regarding star-galaxy separation.  Database and software resources used in
this paper include the SDSS Catalog Archive Server Jobs (CASJobs) System, the
UKIDSS WFCAM Science Archive (WSA), the VizieR catalogue service, the IDL
Astronomy User's Library, \textsc{idlutils}, Astromatic software, and Starlink
Tables Infrastructure Library Tool Set (STILTS).  I.~Baldry acknowledges
funding from Science and Technology Facilities Council (STFC) and Higher
Education Funding Council for England (HEFCE).

\setlength{\bibhang}{2.0em}
\setlength\labelwidth{0.0em}



\label{lastpage}

\end{document}